\documentclass[12pt]{article}
\pdfoutput=1

\usepackage{amsmath}
\usepackage{amsfonts}
\usepackage{amssymb}

\usepackage{hyperref}
\newlength{\abstractwidth}
\usepackage{color}
\usepackage{graphicx}

\renewcommand{\thefootnote}{\fnsymbol{footnote}}
\renewcommand{\thanks}[1]{\footnote{#1}}
\newcommand{\starttext}{
\setcounter{footnote}{0}
\renewcommand{\thefootnote}{\arabic{footnote}}}

\newcommand{\bea}{\begin{eqnarray}}
\newcommand{\eea}{\end{eqnarray}}
\newcommand{\ee}{\end{equation}}
\newcommand{\be}{\begin{equation}}

\newcommand{\no}{\nonumber}

\def\cN{{\cal N}}

\def\Re{{\rm Re}}
\def\Im{{\rm Im}}
\def\tr{{\rm tr}}
\def\det{{\rm det}}

\def\half{ {1\over 2}}
\def\p{\partial}

\def\no{\nonumber}

\long\def\symbolfootnote[#1]#2{\begingroup%
\def\thefootnote{\fnsymbol{footnote}}\footnote[#1]{#2}\endgroup}

\begin{document}
\starttext
\setcounter{footnote}{0}


\bigskip

\begin{center}

{\Large \bf  Boundary entropy of supersymmetric Janus solutions }

\medskip

\vskip .4in

{\large  Marco Chiodaroli$^{a}$, Michael Gutperle$^{a}$, Ling-Yan Hung$^{b}$ }
\vskip .2in

{$\ ^{a}$ \sl Department of Physics and Astronomy }\\
{\sl University of California, Los Angeles, CA 90095, USA}\\
{\tt \small mchiodar@ucla.edu; gutperle@physics.ucla.edu}

\vskip .3in

{$\ ^{b}$\sl Perimeter Institute, Waterloo, Ontario N2L 2Y5,  Canada}\\
{\tt  \small jhung@perimeterinstitute.ca}
\end{center}

\vskip .2in

\begin{abstract}

\vskip 0.1in

In this paper we compute the  holographic boundary entropy for  half-BPS Janus deformations of the   $AdS_3\times S^3\times T^4$
vacuum of  type IIB supergravity.
Previous work \cite{Chiodaroli:2009yw} has shown that there are two independent deformations of this sort. In one case, the six-dimensional dilaton
jumps across the interface, while the other case displays a jump of axion and four-form potential.

In case of a jump of the six-dimensional dilaton, it is possible to compare the holographic result with
the weak-coupling result for a two-dimensional interface CFT where the radii of the compactified bosons jump across the interface.
We find exact agreement between holographic and CFT results.
This is to be contrasted with the holographic calculation for the non-supersymmetric Janus solution,
which agrees with the CFT result only at the leading order in the jump parameter.

We also examine the implications of the holographic calculation in case of a solution with a jump in the axion, which can
be associated with a deformation of the CFT by the $Z_2$-orbifold twist operator.

\end{abstract}

\baselineskip=16pt
\setcounter{equation}{0}
\setcounter{footnote}{0}

\newpage



\baselineskip 16pt

\section{Introduction}
\setcounter{equation}{0}

Conformal field theories with boundaries or interfaces are the object of many interesting applications.
A conformal boundary in a $d$-dimensional conformal field theory is a co-dimension one surface
that is invariant under $d-1$-dimensional conformal transformations.
In condensed matter physics, these theories are used to describe impurities in critical systems.
Cardy  \cite{Cardy:1989ir} initiated the project of classifying  all conformal boundary conditions for
two-dimensional CFTs.
In string theory, boundary conformal field theories are employed for the world-sheet description of D-branes.

Conformal interfaces provide a generalization of boundary CFTs.
In an interface theory, two different conformal theories, $CFT_1$ and $CFT_2$, are separated by a hypersurface of co-dimension one.
The folding trick \cite{Oshikawa:1996dj,Bachas:2001vj} relates a two-dimensional conformal interface theory to a boundary CFT
in the tensor product $CFT_1\otimes CFT_2$.
Consequently, the folding trick can be employed to classify possible  interface theories, calculate reflection and transmission coefficients
for bulk excitations, study bulk and boundary perturbations and renormalization group flows.

In this paper,  we will examine the boundary entropy, which can be obtained from  the ground state degeneracy, or g-factor, of the boundary CFT
 \cite{Affleck:1991tk}.
The boundary entropy  of the folded theory can be interpreted as the entropy associated with the interface. This quantity
is  universal  and constitutes the  analogue of the central charge for a boundary CFT.
The boundary entropy can help  classify conformal boundary conditions and give information about the low energy spectrum of
the system.
In fact, this quantity is measurable in experiments, as demonstrated by studies in candidate materials exhibiting
the Kondo-effect \cite{PhysRevLett.67.2882}. In \cite{Calabrese:2004eu} it was further argued that the boundary entropy is related to the
finite part of the entanglement entropy.

The AdS/CFT correspondence
\cite{Maldacena:1997re,Gubser:1998bc,Witten:1998qj} is a  powerful tool  for  studying  conformal field theories employing
 dual gravitational theories in Anti de-Sitter spacetimes. In particular, the construction of defects and interfaces
in the probe approximation was initiated in \cite{Bachas:2001vj,Karch:2000gx,Aharony:2003qf},
with the analysis of branes spanning $AdS_{d}$ submanifolds inside an $AdS_{d+1}$ space.

Moreover, the so-called Janus-solution was constructed in  \cite{Bak:2003jk}. This solution is  a fully  back-reacted
solution of type IIB supergravity that is locally asymptotic to $AdS_5\times S^5$.
The Janus solution is the holographic dual of an interface theory in which the gauge
coupling is constant throughout the bulk of two 3+1-dimensional half-spaces, but is allowed to jump across
a planar 2+1-dimensional interface, where the half-spaces are glued together.

The literature examines many generalizations of the original Janus solution
\cite{Clark:2005te,Lunin:2006xr,Lunin:2007ab,Yamaguchi:2006te,Gomis:2006sb,Gomis:2006cu,D'Hoker:2007xy,D'Hoker:2007xz,D'Hoker:2008wc,D'Hoker:2008qm,D'Hoker:2009my},
including solutions where the interface preserves up to one half of the supersymmetries of the bulk theory.

In the present paper, we study  supersymmetric   solutions of type IIB supergravity
which are locally asymptotic to $AdS_3\times S^3\times T^4$.
These solutions were first constructed in \cite{Chiodaroli:2009yw}
\footnote{For earlier work in this direction see \cite{Kumar:2002wc,Kumar:2003xi,Kumar:2004me}.}, and are the holographic duals
of various marginal deformations of two-dimensional $\mathcal{N}=(4,4)$ super-conformal field theory.

A prescription for the holographic calculation of the entanglement entropy of a conformal field theory was given in \cite{Ryu:2006bv,Ryu:2006ef}.
In  \cite{Azeyanagi:2007qj} this holographic prescription was used to calculate boundary entropy for probe branes and for
a non-supersymmetric Janus solution with $AdS_3 \times S^3 $ asymptotics.

The goal of the present paper is to calculate the boundary entropy holographically for the half-BPS Janus solution
of \cite{Chiodaroli:2009yw} and compare the result with a weak-coupling calculation in the dual two-dimensional CFT.
In particular, we specialize to the case where only $D1$- and $D5$-brane charges are present.
In this case, the Janus solution describes an interface where two operators in the CFT corresponding to the six dimensional dilaton  and to the linear combination of the axion and  RR four form.

The main result of this paper is the exact agreement between the two calculations in the case in which the interface CFT
displays only a jump of the $T^4$ volume across the interface.
This is in contrast to the non-supersymmetric case where the calculations only agree to the leading order in the deformation parameter.
While we have not performed a rigorous calculation in the presence of  $Z_2$ orbifold twist operator  deformation jump on the CFT side, we have a number of observations, as explained in Section \ref{sec5}, that suggest complete agreement with the supergravity result also in this case.

The organization of the paper is as follows. In Section \ref{sec1}, we briefly review the boundary entropy and
its relation to the entanglement entropy.  In
addition, we give a brief review of the prescription for the holographic calculation of the entanglement entropy
discussed in \cite{Ryu:2006bv,Ryu:2006ef}.
In Section \ref{sec2}, we discuss the regularization  near the boundary of the $AdS_3$ space which will be important for the calculation of
 the boundary entropy.
 In Section \ref{sec3}, we review the non-supersymmetric $AdS_3$  Janus solution found in \cite{Bak:2007jm} and the holographic calculation of the interface
entropy performed in \cite{Azeyanagi:2007qj}. We pay particular attention to the regularization of the holographic entropy function.
In Section \ref{sec4}, we generalize the framework for the holographic computation of the entanglement entropy introduced
in \cite{Ryu:2006bv,Ryu:2006ef} to the case of a spacetime with geometry of the form $AdS_p \times S^q \times \Sigma$, where $\Sigma$ is a two-dimensional surface.
We examine the supersymmetric Janus solution obtained in \cite{Chiodaroli:2009yw}, and compute the holographic entanglement entropy as a function
of the deformation parameters.
In Section \ref{sec5}, we perform the  CFT calculation of the boundary entropy in case of a jump in the radii of the compact bosons
and find exact agreement with the result from Section \ref{sec4}. Moreover, we explore the implications of our result on the properties of
correlators of $Z_2$ twist fields.
We conclude the paper with a discussion of our results.

\section{Boundary and entanglement entropy}\label{sec1}
\setcounter{equation}{0}
The logarithm of the partition function for a two-dimensional conformal
theory defined on a spatial segment of length $L$ is given by
\be
 \log Z\sim   \log  \tr(e^{-\beta H_{AB}}) = \log(g_A g_B) + {c \pi \over 6 \beta}L
\ee
This expression is valid in the limit $L>> \beta$.
Here $H_{AB}$ is the open string Hamiltonian associated with the conformal boundary conditions on the two ends of the strip,
denoted by $A$ and $B$.
The universal factor $g_A$ is interpreted as the ground state degeneracy, or g-factor, associated with the conformal boundary $A$
\cite{Affleck:1991tk}. A modular transformation relates the open string annulus partition function
to a closed string cylinder partition function,
\be
Z= \langle A\mid e^{-L H_{cl}} \mid B\rangle
\ee
where $\mid A\rangle$ is the boundary state which realizes the boundary condition $A$ on the closed string channel CFT. In the limit $L/\beta\to \infty$,
it can be shown that the $g_A$ factor is given by
\be\label{gfacta}
g_A=\langle 0 \mid A\rangle
\ee
The logarithm of $g_A$ is called the boundary entropy and counts the ground state degeneracy of the boundary theory.

It has been argued in  \cite{Calabrese:2004eu} (see also \cite{Azeyanagi:2007qj} for a recent discussion)
that the boundary entropy can be associated with the entanglement entropy in a system with a boundary.
Consider a CFT defined on a half-line with a conformal boundary condition  $A$ at $x=0$.
The system can be divided into a subsystem ${\cal A}$, defined on the interval $[0,l]$, and its complement  ${\cal B}$, defined on  $[l,\infty]$.
The total space of states is given by the product $H=H_{\cal A}\otimes H_{\cal B}$.
A reduced density matrix can be defined by tracing over all states in ${\cal B}$,
\be
\rho_{\cal A} = \tr_{H_{\cal B}} \rho
\ee
where $\rho$ is the density matrix of the total system (at zero temperature this is just the projector on the ground state).
The entanglement entropy is then defined as
\bea\label{entangent}
S_{\cal A} =- \tr_{H_{\cal A}} \rho_{\cal A} \log \rho_{\cal A}
\eea
The entanglement entropy takes the following form in the limit of large $l$  \cite{Calabrese:2004eu},
\be
S_{\cal A} = {c\over 6} \log {l\over a} + \log g_A
\ee
where $a$ is a UV cutoff and $g_A$ is precisely the g-factor (\ref{gfacta}) associated with the conformal boundary $A$.\footnote{An additional term $c'$ was included in \cite{Calabrese:2004eu}. This term is non-universal and independent of the
presence of the boundary. We compute $g_A$ as the difference between the entanglement entropy with an interface
and the entropy of the same system  without an interface. Therefore, $c'$ will not contribute to our results.}

A proposal to calculate the entanglement entropy of a $CFT_{d}$ with a dual description
as a gravitational theory in $AdS_{d+1}$ was discussed in \cite{Ryu:2006bv,Ryu:2006ef}.
Working in Poincar\'{e} coordinates, the CFT is defined on Minkowski space $R^{1,d-1}$ which can be
 thought of as the boundary of $AdS_{d+1}$.  We consider a static setup  where we choose a particular time slice on the boundary.

\begin{figure}
\centering
\includegraphics[scale=0.17]{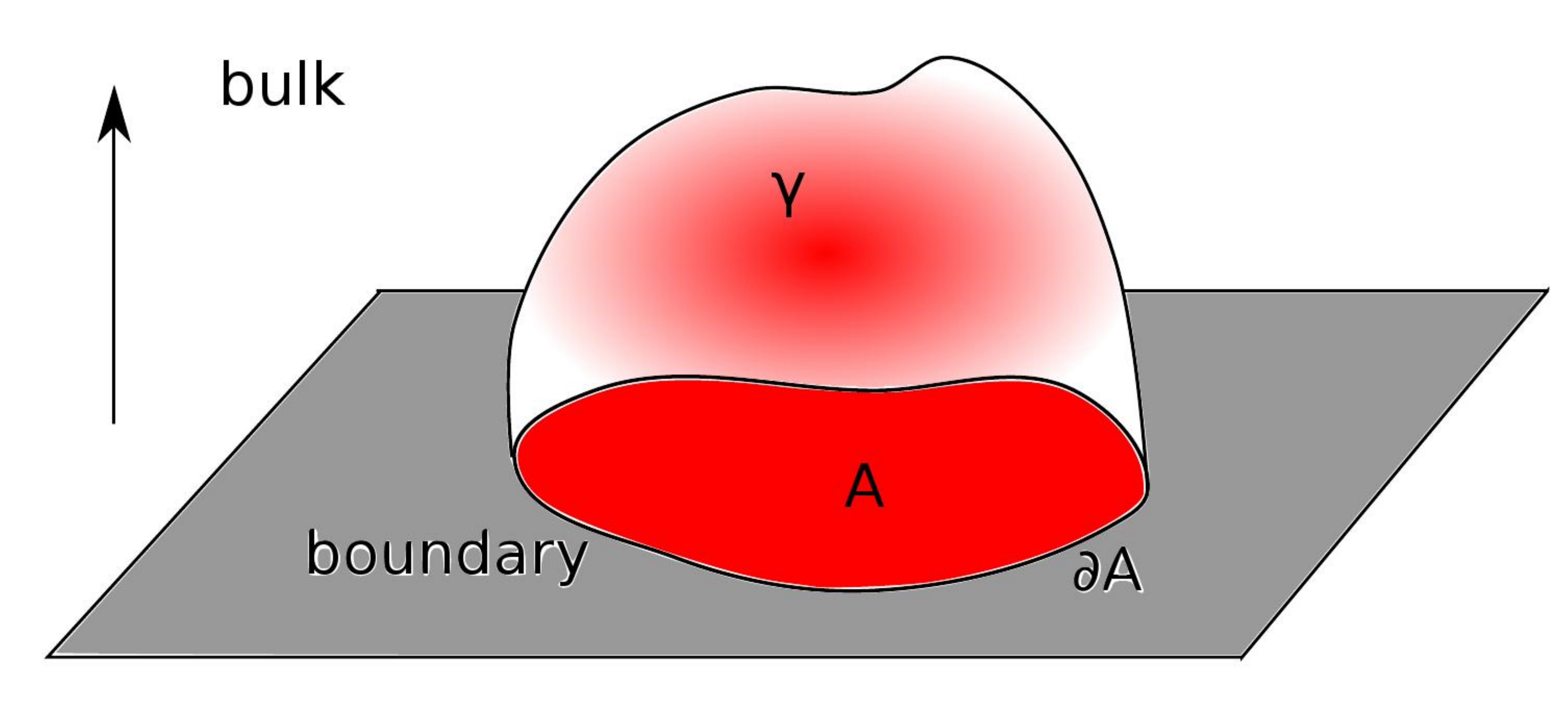}
\caption{ Minimal surface for the holographic calculation of boundary entropy.}
\label{fig-entangle}
\end{figure}

The subsystem ${\cal A}$ is a $d$-dimensional spatial region in the constant-time slice.
The boundary of ${\cal A}$ will be denoted by $\partial {\cal A}$ (see figure \ref{fig-entangle}).

One can find a static minimal surface $\gamma_{\cal A}$ which extends into the $AdS_{d+1}$ bulk and ends on $\partial {\cal A}$
as one approaches the boundary of $AdS_{d+1}$.
The holographic entanglement entropy can then be calculated as follows  \cite{Ryu:2006bv,Ryu:2006ef},
\bea
S_A = {{\rm Area}(\gamma_{\cal A}) \over 4 G^{(d+1)}_{N}}
\eea
where ${\rm Area}(\gamma_{\cal A})$ denotes the area of the minimal surface $\gamma_{\cal A}$ and $ G^{(d+1)}_{N}$ is
the Newton constant for $AdS_{d+1}$ gravity.
In the case of $AdS_3$, the area ${\cal A}$ is an interval and the boundary $\partial {\cal A}$ is a collection of points.
The minimal surface is a spacelike geodesic connecting these points.

\section{Ultraviolet regularization of $AdS$-sliced metrics}
\label{sec2}
\setcounter{equation}{0}

The starting point for the construction of the relevant Janus solutions is the $AdS_2$ slicing of $AdS_3$,  given by
\bea\label{ads3met}
ds^2= R^2_{AdS_3} \Big(dx^2 +\cosh^2 x {dz^2-dt^2\over z^2} \Big)\label{ads-slicing}
\eea
The  structure of the $AdS_3$ boundary  in this coordinates is more complicated than the one encountered with the Poincar\'{e} patch.
In particular, there are three boundary components. Two $1+1$-dimensional half-spaces can be reached by taking the limit $x\to \pm \infty$.
The two half spaces are glued together at a $0+1$-dimensional world-line, which is reached by taking $z\to 0$.
It is useful to perform the change of coordinates,
\be
\zeta= \tanh(x)
\ee
Under this change, the metric (\ref{ads3met}) becomes
\be
ds^2= R^2_{AdS_3} \Big({d\zeta^2 \over (1-\zeta^2)^2} + {1\over 1-\zeta^2 } {dz^2-dt^2\over z^2}\Big)
\ee
In the calculation of the entanglement entropy one obtains expressions which diverge near the boundary of the bulk spacetime.
 In the field theory, this phenomenon can be interpreted as an ultraviolet divergence.
Hence, in order to perform the calculation we have to introduce an ultraviolet regulator.
The introduction of Fefferman-Graham coordinates provides a systematic procedure for regularizing the $AdS$-sliced metrics,
as discussed in Appendix B of \cite{Papadimitriou:2004rz}.

In case of the metric (\ref{ads3met}), the  Fefferman-Graham coordinates are equivalent to the Poincar\'{e} patch,
and can be introduced as
\be\label{fgcoord}
\zeta= {\eta\over \sqrt{\xi^2+\eta^2}}, \quad  z= \sqrt{\xi^2+\eta^2}
\ee
With this change, the metric becomes
\be
ds^2= {R_{AdS_3}^2\over \xi^2} \Big( d\xi^2+ d \eta^2 -dt^2\Big)
\ee
The boundary of the bulk spacetime is now reached by $\xi\to 0$.
The interface is located at $\xi=0, \eta=0$.
 Note that there are subtle issues with the order of limits if one approaches the boundary and the interface at the same time.

The global coordinate change to a Fefferman-Graham system is not known  for the supersymmetric Janus solution
 employed in this paper.
In the following analysis, we will be interested mainly in imposing an ultraviolet cutoff in the region away from the interface,
i.e. where $z>>0$.
Note that a more complete analysis is needed for holographic calculations of bulk/boundary operator expansions in the CFT,
as one is interested in the behavior of the solution as $z\to 0$.

By taking  $x\to \pm \infty$ and keeping $z$ finite, one reaches the boundary staying away from the interface.
In this limit, we will encounter metrics with a slightly more general form than (\ref{ads3met}),
\bea\label{boundlima}
\lim_{x\to \pm \infty} ds^2= R_{AdS_3}^2\Big(  dx^2 +{\lambda_{\pm}\over 4}  e^{\pm 2x} {dz^2-dt^2\over z^2}\Big) + o(1)
\eea
One can absorb the constant $\lambda_\pm $ by a shift in $x$,
\be\label{shiftx}
x=\tilde x  \mp {1\over 2} \log (\lambda_{\pm} )
\ee
and one gets
\bea
\lim_{x\to \pm \infty} ds^2= R_{AdS_3}^2\Big(  d\tilde x^2 + {1\over 4}e^{\pm 2\tilde x} {dz^2-dt^2\over z^2}\Big) + o(1)
\eea
In this limit, the Fefferman-Graham coordinate change  becomes
\bea\label{fgchangeb}
\tilde x\to +\infty, \quad \xi\to 0, \quad \eta>0 &:& \quad \quad e^{-2\tilde x} = {1\over 4} {\xi^2 \over \eta^2}, \quad  z= \eta \Big(1+ {1\over 2} {\xi^2\over \eta^2} \Big)\no\\
\tilde x\to -\infty,\quad  \xi\to 0,\quad  \eta<0 &:& \quad \quad e^{2\tilde x} = {1\over 4} {\xi^2 \over \eta^2}, \quad  z= |\eta| \Big(1+ {1\over 2} {\xi^2\over \eta^2} \Big)\no\\
\eea
The above equations were  obtained using the relations
\bea
\tilde x\to +\infty, \quad \quad \zeta= 1-2 e^{-2\tilde x}+ \cdots \no\\
 \tilde x\to -\infty, \quad \quad \zeta= -1+2 e^{2\tilde x}+ \cdots
\eea
In both cases, the metric becomes
\be
\lim_{|x| \to \pm \infty} ds^2= {R_{AdS_3}^2\over \xi^2} \Big( d\xi^2+ d \eta^2 -dt^2\Big)+\cdots
\ee
The interface is located at $\eta=0$, and in the two cases the coordinate $\eta$ is restricted to $\eta>0$ and $\eta<0$ respectively.
Note that near $\eta=0$  the change of coordinates is more complicated.
In particular, the coordinate change (\ref{fgchangeb}) appears not to be smooth at $\eta=0$.
However, this change of coordinates
is valid as long as  one is not approaching the interface while approaching  the boundary.
If we now set the cutoff  at $\xi=\epsilon$  and consider a point $|\eta|=z_0$, we can use (\ref{shiftx}) and (\ref{fgchangeb})
to obtain a relation between the cutoff $\epsilon$ and $x$ in the two asymptotic regions,
\bea\label{cutoffa}
x\to +\infty&:& \quad \epsilon= {2 z_0 \over \sqrt{\lambda_+} }e^{-x_\infty}\no\\
x\to -\infty&:& \quad \epsilon= {2 z_0 \over \sqrt{\lambda_-} }e^{x_{-\infty}}
\eea

\section{Non-supersymmetric Janus solution}\label{sec3}
\setcounter{equation}{0}

The boundary entropy for a non-supersymmetric Janus solution with $AdS_3$ asymptotics was calculated in \cite{Azeyanagi:2007qj}.
 We reproduce the result of  \cite{Azeyanagi:2007qj} by introducing an ultraviolet regulator near the $AdS_3$ boundary.

The non-BPS Janus deformation of the $AdS_3\times S^3\times T^4$ vacuum of type IIB supergravity was found in \cite{Bak:2007jm}.
The ten-dimensional metric is given by
\bea\label{nonbpsjanusa}
ds^2= e^{\phi\over 2} \big(ds_{3}^2+ ds_{S^3}^2\big)+ e^{-\phi/2} ds_{M_4}^2
\eea
where the three-dimensional metric $ds_{3}^2$ is
\bea
ds_3^2= R_{AdS_3}^2 \Big( dx^2+ f(x) ds_{AdS_2}^2\Big)
\eea
the function $f(x)$ has the following expression,
\bea
f(x)&=& {1\over 2} (1+ \sqrt{1-2 \gamma^2} \cosh(2 x)\Big)
\eea
The dilaton is given by
\bea
\phi(x)&= &\phi_0 +{1\over \sqrt{2}} \log \left({ 1+ \sqrt{1-2 \gamma^2}  +\sqrt{2} \gamma \tanh x\over
 1+ \sqrt{1-2 \gamma^2}  -\sqrt{2} \gamma \tanh x}\right)
\eea
Note that if we consider the compactification of the ten-dimensional theory to six-dimensions,
the resulting action is not in the Einstein frame.
To bring the six-dimensional action to the Einstein frame we have to multiply the six dimensional part of the metric
by $e^{-\phi/2}$,
\be
ds_{6,E}^2=  R_{AdS_3}^2 \big( dx^2+ f(x) ds_{AdS_2}^2\big) + R_{AdS_3}^2 \big(dy^2+ \sin^2y \; ds_{S^2}^2 \big)
\ee
The expansion of $f(x)$ as $x\to \pm \infty$ is
\bea
\lim_{x\to \pm \infty} f(x)= {1\over 4} \sqrt{1-2\gamma^2} e^{\pm 2x} +o(1)
\eea
Comparison with (\ref{boundlima}) shows that in this case the constants $\lambda_\pm$ are given by
\bea
\lambda_{\pm}= \sqrt{1-2\gamma^2}
\eea

\subsection{Holographic boundary entropy}
\begin{figure}
\centering
\includegraphics[scale=0.54]{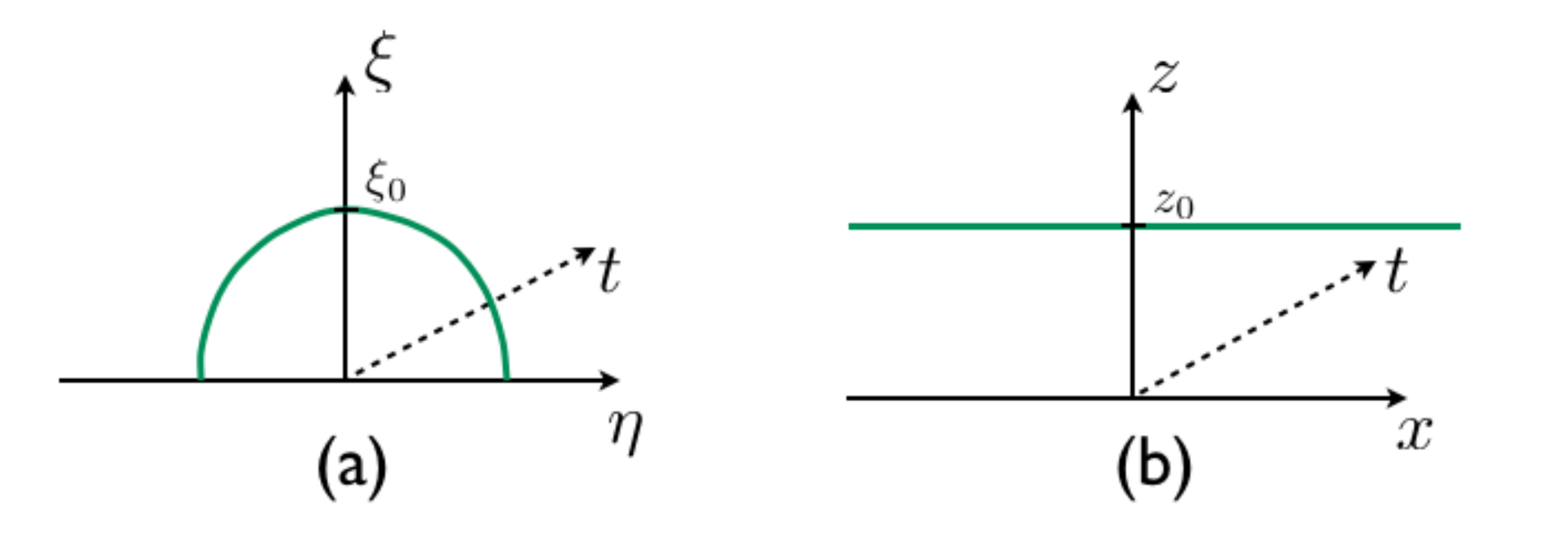}
\caption{(a) Minimal surface for the holographic entanglement entropy in Poincare coordinates (b) Minimal surface in the $AdS_2$ slicing of $AdS_3$. }
\label{fig:3}
\end{figure}
The geodesic which was used in \cite{Azeyanagi:2007qj} to compute the entanglement entropy has a particularly simple form:
the $z$ coordinate stays constant,
\be z =z_0 \ee
while $x$ varies from $- \infty$ to $+ \infty$. This corresponds to a symmetric region around the interface as depicted in Figure \ref{fig:3}.

With this choice, the subsystem $\cal A$ is a symmetric interval around the interface given by $[-z_0,z_0]$.
In this particular case, the geodesic length is given by
\be\label{geolength}
\Gamma(\gamma) = R_{AdS_3}\int dx =  R_{AdS_3}\big(x_{\infty} (\gamma) - x_{-\infty} (\gamma)\big)
 \ee
where $x_{\pm \infty}$ is the $x$ coordinate evaluated at the cutoff.
We can now use the relation between $x_{\pm \infty}$ and the cutoff $\epsilon$,
\bea
x_\infty &=& -\log \epsilon  -{1\over 2} \log \lambda_+ + \log(2z_0)\no\\
x_{-\infty} &=& \log \epsilon  +{1\over 2} \log \lambda_- - \log(2z_0)
\eea
Hence
\bea
\Gamma(\gamma)/R_{AdS_3} &=& x_{\infty} (\gamma) - x_{-\infty} (\gamma)  \no\\
&=& -2 \log \epsilon -{1\over 2} (\log \lambda_+ +\log \lambda_-) + 2 \log (2z_0)\no\\
&=& -2 \log \epsilon -\log (\sqrt{1-2\gamma^2})+ 2 \log (2z_0)
\eea
 The entanglement  entropy is then given by  the difference between the geodesic length in the Janus geometry
and the length in $AdS_3$ evaluated at the same value of the cutoff $\epsilon$.
The difference of the two lengths is then finite and independent of the cutoff,
 \bea\label{karchresult}
 S_{bdy}&=& {\Gamma(\gamma)-\Gamma(0) \over G_3}  = - {R_{AdS_3}\over  4 G_3}  \log  (\sqrt{1-2\gamma^2}) =  N_{D1} N_{D5}  \log {1\over \sqrt{1-2\gamma^2}}
 \eea
We used  the Brown-Henneaux formula \cite{Brown:1986nw}  for   the central charge of the the CFT,
 \bea
 {3 R_{AdS_3} \over 2 G_3} = c= 6 N_{D1} N_{D5} \label{rel-central}
 \eea
$N_{D1/D5}$ are the number of $D1$- and $D5$-branes which realize the CFT.
With equation (\ref{karchresult}) we have reproduced the result for the boundary entropy (\ref{karchresult}) found in  \cite{Azeyanagi:2007qj}.

\section{Supersymmetric Janus solution}\label{sec4}
\setcounter{equation}{0}
In this section we will calculate the holographic boundary entropy for the supersymmetric Janus solution found  in \cite{Chiodaroli:2009yw}.
More details about the solution can be found in Appendix \ref{appa}.

We consider type IIB supergravity compactified on $ M_4$, where $M_4$ is either $K_3$ or $T^4$.
In the present paper we will use a four-torus since the dual CFT is simpler.

The ten-dimensional metric ansatz is a fibration of $AdS_2 \times S^2 \times M_4$ over a two dimensional Riemann surface $\Sigma$
\bea\label{bpsmeta}
ds^{2} &=& f_{1}^{2 } ds^{2}_{AdS_{2}} + f^{2}_{2}ds^{2}_{S^{2}} + f^{2}_{3}ds^{2}_{M_4}  + \rho^{2 }dz  d\bar z
\eea
All fields will depend on the coordinates $z,\bar z$ of  $\Sigma$.  For the supersymmetric
Janus solution, $\Sigma$ is an infinite  strip  with coordinates
\bea
w=x+i y, \quad \quad  x\in {}[-\infty,+\infty{} ], \;\; y\in {}[0,\pi{}]
\eea
The boundaries of the strip  are at $y=0,\pi$.
The supersymmetric Janus solution depends on four parameters $k,L,\theta$ and $\psi$.
$\psi$ parameterizes the jump of the six-dimensional dilaton across the interface
and $\theta$ parameterizes the jump of the axion \footnote{ A constant $c$ which appeared in \cite{Chiodaroli:2009yw} has been set to one. Moreover,
in the previous paper we used the scalar $\Phi$ which is related to the standard dilaton by $\phi=-2\Phi$.}. Note that setting  $\psi=0$ and $\theta=0$ gives the $AdS_3\times S^3$ vacuum.
\begin{figure}
\centering
\includegraphics[scale=0.40]{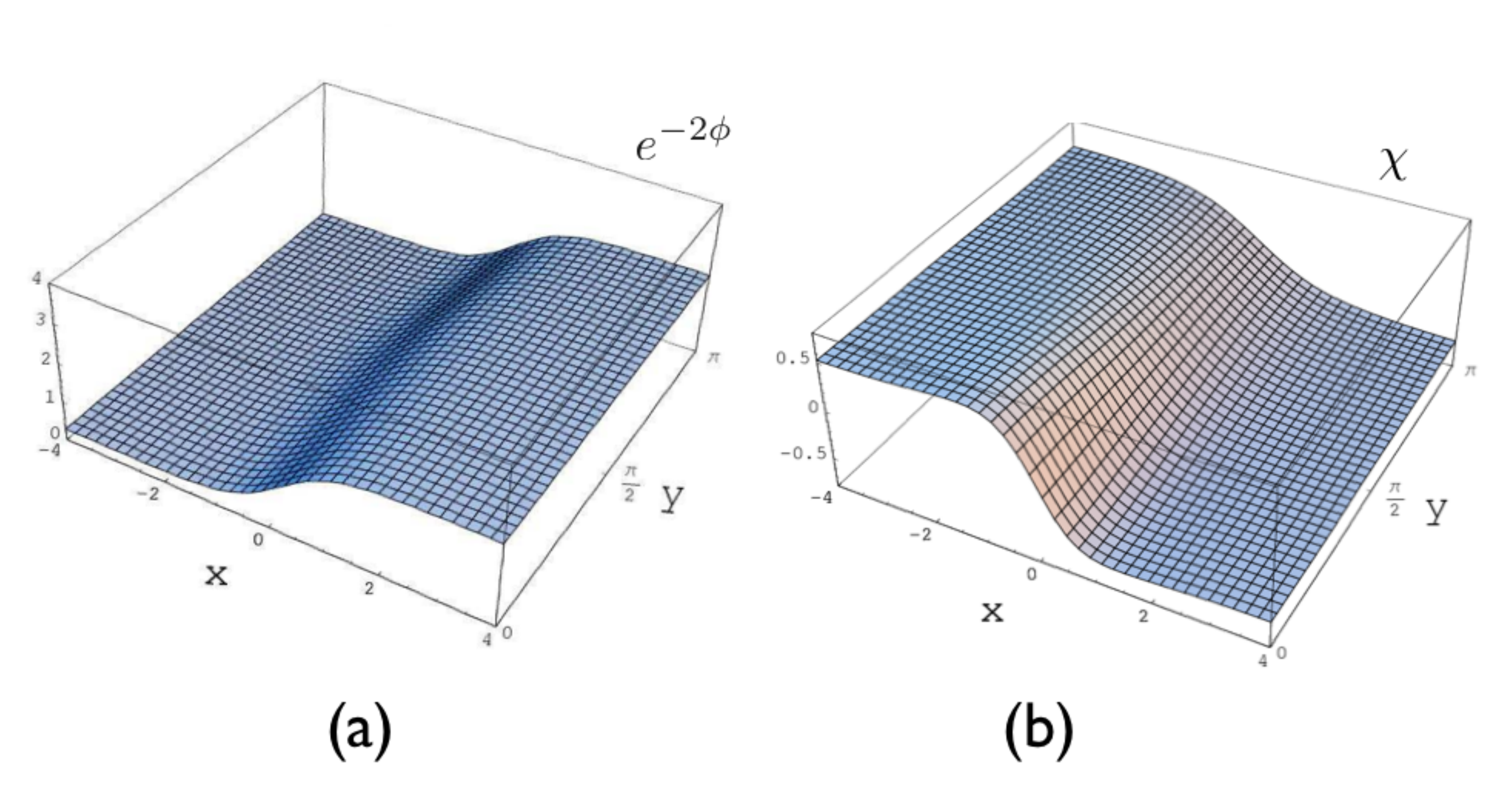}
\caption{ (a) Plot of the dilaton $e^{-2\phi}$ for the deformation $\theta=0, \psi=\half$ depicting the jump in the dilaton
as $x\to \pm \infty$. (b) Plot of the axion $\chi$ for the solution  $\psi=0,\theta=\half$
 depicting the jump in the axion as $x\to \pm \infty$.}
\label{fig-psi}
\end{figure}
The dilaton and axion are given by
\bea
 e^{-2\phi} &=&  k^4 {\cosh^2 (x + \psi)  {\rm sech}^2 \psi + \big(  \cosh^2 \theta -  {\rm sech}^2  \psi \big) \sin^2 y \over \big( \cosh  x  - \cos y \tanh \theta \big)^2} \\
\chi &=& - {k^2 \over 2}  {  \sinh 2 \theta  \sinh x - 2 \tanh \psi \cos y \over \cosh x \cosh \theta - \cos y \sinh \theta} \eea
The metric factors for $\Sigma$ and $T^4$ are given by
\bea
\rho^4&=& { e^{- \phi} } {L^2\over k^2}  { \cosh^2 x \cosh^2 \theta - \cos^2 y \sinh^2 \theta \over \cosh^2 (x + \psi) } \cosh^4 \psi \no\\
f_3^4 &=&e^{- \phi} {4\over k^2} {\cosh x \cosh\theta-\cos y \sinh \theta \over \cosh x \cosh\theta+\cos y \sinh \theta  }
\eea
The following expressions for the  $AdS_2$ and $S^2$ metric factors will be useful,
\bea
{f^2_1 \over \rho^2} &=& {\cosh^2 \big(x + \psi \big)\over \cosh^2 \theta \cosh^2 \psi } \no \\
{\rho^2\over f_2^2}  &=&{1\over \sin^2 y} + {\cosh^2 \theta \cosh^2 \psi - 1 \over  \cosh^2 \big(x + \psi \big) } \label{relmetr3}
\eea
The Page charges, defined in Appendix \ref{appa}, can be calculated for the present solution.
 One finds that  the fundamental string and $NS5$-brane charges vanish, i.e. $Q_{F1}=0$ and $Q_{NS5}=0$. The $D1$- and $D5$-brane
charges are given by
\bea
 Q_{D5}&=& 4 \pi^2 k L \cosh \psi \;  \cosh \theta \no\\
 Q_{D1}&=& {16 \pi^2 L\over k} \cosh \psi \;  \cosh \theta
 \eea
The dual CFT is a $\mathcal{N}=(4,4)$ superconformal field theory which can be understood, at a particular point of its moduli space,
as the $(M^4)^{Q_{D1} Q_{D5}} / S_{Q_{D1} Q_{D5}}$ orbifold  sigma-model.
 The central charge of the CFT can be expressed in terms of the charges as follows
\bea\label{centralc}
c&=& {6\over 4 \pi  k_{10}^2} Q_{D1}Q_{D5}  = {3 \times 32 \; \pi^3 L^2 \over k_{10}^2} \cosh^2\psi\;  \cosh^2\theta \eea

The limit $x \to \pm \infty$ corresponds to approaching the $AdS$ boundary, where the scalar fields behave as follows
\bea\label{radii}
\lim_{x\to \pm \infty} e^{-\phi}f_3^4 &=& 4   k^2 {e^{\mp 2 \psi}\over \cosh^2 \psi}
\eea
This combination of scalars is precisely the six-dimensional dilaton,
and is dual to the volume of the four-torus of the  orbifold CFT.
Hence, the deformation corresponds to an interface CFT where the volume jumps across the interface.

Similarly, the combination
\bea \label{twistcoup}
\lim_{x\to \pm \infty} {e^{\phi \over 2 }  f_3^2} \chi - 4 {e^{-{\phi \over 2} } \over f_3^2} C_4 &=& \pm 4 k  \sinh \theta
\eea
is  dual to the $Z_2$  orbifold twist operator of the $(M_4)^N /S_N$ orbifold.
Solutions with non-zero $\theta$ correspond
to interfaces in which the orbifold CFTs on each side are at two different points in their moduli space.

  \subsection{Calculation of the holographic boundary entropy}\label{calchent}

The calculation of the holographic boundary entropy for the non-supersymmetric Janus solution was
simplified by the fact the three-sphere was left undeformed.
The calculation of the minimal surface could be done by compactifying  the ten-dimensional theory
on $S^3$ and $T^4$ obtaining an effective three-dimensional theory.
This reduction is equivalent to considering a minimal surface wrapping these spaces,
in accordance with the principle that the entanglement entropy is a trace over all states. 

Supersymmetric Janus solution in various dimensions have a more complicated structure. In general, their geometry is given by an
$AdS_p \times S^q$ fibration over a Riemann surface $\Sigma$. In analogy with the simpler case of non-supersymmetric solutions,
we propose to compute the entanglement entropy using a $p+q$-dimensional minimal surface which spans
the sphere $S^q$ as well as the Riemann surface $\Sigma$.

It is useful to consider an intermediate step and compactify the metric (\ref{bpsmeta}) along the compact directions.
In order to have the correct Einstein-Hilbert action, we rescale by a factor of $f_3^{-2}$
to obtain the resulting metric in the six-dimensional  Einstein frame,
 \bea\label{sixmetric}
ds^2_{6,E} ={\rho^2  f_3^2}\Big( {\cosh^2 (x + \psi) \over \cosh^2 \psi \cosh^2 \theta} {dz^2- dt^2 \over z^2}  + dx^2 + dy^2 \Big) + {f^2_2  f_3^2} ds^2_{S^2}
\eea
We now use the following expressions for the limit $x\to \pm \infty$,
\bea
\lim_{x\to \pm \infty} e^{-\phi} &=&  k^2 e^{\pm \psi} {\rm sech} \psi \no\\
\lim_{x\to \pm \infty}  \rho^2 &=& L\;  e^{\mp \psi/2} \cosh  \theta \cosh^{3/2}\psi\no\\
\lim_{x\to \pm \infty}  f_3^2  &=&  2 {e^{\pm \psi/2} \over \cosh^{1/2} \psi}\no\\
\lim_{x\to \pm \infty}  f_2^2 &=& \rho^2 \sin^2y = L  \;e^{\mp \psi/2}  \cosh  \theta \cosh^{3/2}\psi\sin^2 y
\eea
Hence the metric (\ref{sixmetric}) becomes
\bea
\lim_{x\to \pm \infty}  ds^2_{6,E} &=&  2 L \; \cosh \theta \cosh \psi \Big( { e^{\pm 2\psi }\over \cosh^2 \psi \cosh ^2\theta} {e^{\pm 2 x}\over 4} {dz^2-dt^2 \over z^2} + dx^2 + dy^2 + \sin^2 y ds_{S_2}^2\Big) \no\\
\eea
We can then identify
\bea
 R_{AdS_3}^2 &=&2L \;\cosh \theta \cosh \psi
 \eea
Note that the $AdS$ radius  is the same on both sides of the interface.
The constants $\lambda_{\pm}$ are then given by
 \bea
 \lambda_{\pm}&=& { e^{\pm 2\psi }\over \cosh^2 \psi \cosh ^2\theta}
\eea
This expression gives us the relevant formula for the cutoff.
We consider the four-dimensional minimal surface located at $z=z_0, \; t=t_0$ and spanning $x, \; y$ and the $ S_2$ directions.
Using the six-dimensional metric (\ref{sixmetric}) we obtain
\bea
A(\theta, \psi)&=& \int d\Omega_2 \int dx \int  dy \;   (f_2^2 f_3^2) \times (\rho^2 f_3^2) \no\\
&=& V_{S^2} \int dx dy {H ^2\rho^2 \over f^2_1} \no \\
&=& 4 L^2 \cosh^2 \psi \cosh^2 \theta  V_{S^2}  \int dy  \sin^2 y \int dx \no\\
&=&  4 L^2  \cosh^2 \psi \cosh^2 \theta  V_{S^3}  \int dx  \no\\
&=& R_{AdS_3}^4 V_{S^3}  \int dx
\eea
We need to compare this area to the one of a minimal surface in an $AdS_3\times S^3$ space with the same curvature radius.
Instead of (\ref{sixmetric}), we have the metric
\be
ds^2_{AdS_3 \times S^3}  = R_{AdS_3}^2\Big( {e^{\pm 2 x}\over 4} {dz^2-dt^2 \over z^2} + dx^2 + dy^2 + \sin^2 y ds_{S_2}^2\Big)
\ee
which is the $AdS_3 \times S^3$ metric with the same cosmological constant as (\ref{sixmetric}). The area in $AdS_3\times S^3$ is
\bea
A_0&=& R_{AdS_3}^4  \int d\Omega_2 \int dx \int  dy \sin^2 y = R_{AdS_3}^4  V_{S_3} \int dx
\eea
and the difference is
\bea
A(\theta, \psi) -A_0 =  R_{AdS_3}^4  V_{S^3} \Big(\Gamma(\theta, \psi) -\Gamma(0)\Big)
\eea
where the geodesic length $\Gamma$ was defined in (\ref{geolength}).
We can now apply the cutoff prescription discussed in Section 3 and get
\bea
A(\theta, \psi) -A_0 &= & -R_{AdS_3}^4   V_{S^3}  {1\over 2}  \big( \log \lambda_+ +\log \lambda_-\big) = 2 R_{AdS_3}^4 V_{S^3}  \log \big( \cosh \theta \cosh\psi\big)
\eea
giving the boundary entropy
\bea
S_{bdy}&=& {A(\theta, \psi) -A_0 \over  4 G_6}\no\\
&=&  {16 \pi^2 L^2 \over  4 G_6}  \cosh^2 \psi \cosh ^2 \theta   \log \big( \cosh \theta \cosh\psi\big) \no\\
&=& {32 \pi^3 L^2 \over  k_{10}^2}  \cosh^2 \psi \cosh ^2 \theta   \log \big( \cosh \theta \cosh\psi\big)
\eea
where we used the relation between the Newton's constant and $k_{10}$,
\be
{1\over 16 \pi G_N} = {1\over 2 k_{10}^2}
\ee
In particular, setting $\theta=0$ gives
\bea\label{holores}
S_{bdy}= {32 \pi^3 L^2 \over  k_{10}^2}  \cosh^2 \psi   \log  (\cosh\psi)
\eea
This expression gives the boundary entropy for a Janus solution in which only the six-dimensional dilaton
jumps across the interface.

\section{CFT calculation of the boundary entropy}\label{sec5}
\setcounter{equation}{0}
In this section we will  discuss the calculation of the boundary entropy on the CFT side.
Preservation of one half of the supersymmetries imposes specific boundary conditions on the fields at the interface.
We use results in the literature to calculate the boundary entropy for the case where only the coupling constant jumps across the interface.

\subsection{Supersymmetric boundary conditions}\label{susyc}

We now discuss the supersymmetries which are preserved in the presence of an interface in two-dimensional $\cN=(4,4)$ super conformal field theory.
For simplicity, we will focus on the free field limit, where the target space is simply the orbifold $(T^4)^N/S_N$.
The free field action of the theory is given in \cite{David:2002wn}, and  can be written as follows in terms of the complex fields defined there,
\bea
S=&& \frac{1}{2}\int d\tau d\sigma \frac{1}{2}(\partial_{+} X^I\partial_{-}{X^I}^\dagger+\partial_{-}X^I\partial_{+}{X^I}^\dagger)\no\\
 &&-( \Psi^I\partial_{-} {\Psi^I}^\dagger-\frac{1}{2}\partial_{-}(\Psi^I{\Psi^I}^\dagger)  + \tilde{\Psi}^I\partial_{+}\tilde{\Psi}^{I\dagger}-\frac{1}{2}\partial_{+}(\tilde{\Psi}^I \tilde{\Psi}^{I\dagger}))
\eea
The fields are functions of $\sigma_\pm = \tau \pm \sigma$, and summations over the indices $I \in \{1,2\}$ are implied.
This action is in  Lorentzian signature.
We also have suppressed the indices $A$ displayed in \cite{David:2002wn}, which run over the $N$ copies of the four-torus.
Fermionic fields with and without a tilde correspond to the right and left-movers respectively.

The action is invariant up to boundary terms under the four supersymmetric transformations. These terms are given by
\bea
&&\delta_{1}S = \int d\tau  \epsilon_1(\Psi^1\partial_-X^{2} - \Psi^{2\dagger}\partial_-X^1 ), \qquad\delta_{1^\dagger}S = \int d\tau \epsilon_{1\dagger}(\Psi^{1\dagger}\partial_-X^{2\dagger} - \Psi^{2}\partial_-X^{1\dagger}),\no\\
&&\delta_{2}S = \int d\tau  \epsilon_2(\Psi^2\partial_-X^{2} - \Psi^{1\dagger}\partial_-X^1 ), \qquad \delta_{2^\dagger}S = \int d\tau  \epsilon_{2\dagger}(\Psi^{2\dagger}\partial_-X^{2\dagger} - \Psi^{1}\partial_-X^{1\dagger} )\no\\
\eea
where the transformations are generated by the four left-moving charges $G^a, G^{a\dagger}, a\in\{1,2\}$. The above expressions are evaluated at the boundary located at $\sigma=\sigma_0$.

The right-moving supercharges $\tilde{G}^a, \tilde{G}^{a\dagger}$ give rise to boundary terms that can be obtained from
the above expressions exchanging $+$ and $-$.

One then has to determine the boundary conditions satisfied by the fields at the interface.
From energy conservation at the boundary it is natural to require that
\be
\left(T(\sigma_+)-\tilde{T}(\sigma_-)\right)\vert_{\sigma_0}=0
\ee
There are however various possible choices for the boundary conditions satisfied
by the supercharges. For $\cN=(4,4)$ supersymmetries,
one can generalize the A- and B-type boundary conditions which are used in the  $\cN=(2,2)$ case.
The boundary conditions satisfied by the R-symmetry current can then be deduced from those of the supercharges.
Generally, if the interface preserves the global $SO(4)$ symmetry, the supercharges have to satisfy
\bea
\left(\begin{array}{c}G^1 \\ G^{2\dagger}\end{array} \right)\pm \left(\begin{array}{c}\tilde{G}^1 \\ \tilde{G}^{2\dagger}\end{array} \right) &=&0, \no \\
\left(\begin{array}{c}G^{1\dagger}\\ G^2\end{array} \right)\pm \left(\begin{array}{c}\tilde{G}^{1\dagger} \\\tilde{G}^2\end{array} \right) &=&0
\eea

The analog of B-type boundary conditions is obtained by taking the minus sign in the above equations. In this
case the $SU(2)_\textrm{L}\times SU(2)_\textrm{R}$ R-currents satisfy the boundary conditions
\be\label{jboundcon}
J^\alpha(\sigma_+)- \tilde{J}^\alpha(\sigma_-) =0, \qquad \alpha \in \{1,2,3\}
\ee

Now we will examine super conformal field theories with two branches joining along the same boundary.
For simplicity,  we will consider only jumps in the radii of the target space torus across the interface.
Using the folding trick as in \cite{Bachas:2001vj},
we can show that this setup is equivalent to considering the direct product of two $\mathcal{N}=(4,4)$ super conformal field theories
with a boundary.

In the free field limit, the boundary conditions satisfied by the bosons would again be given by \cite{Bachas:2001vj,Azeyanagi:2007qj}
\be
\frac{\dot{X}^I_1}{r^I_1}= \frac{\dot{X}^I_2}{r^I_2}\vert_{\sigma=\sigma_0},\qquad \sum_{i=1,2} r^I_i X'^I_i \vert_{\sigma=\sigma_0}=0 \label{bosonicbc}
\ee
Picking the B-type conditions as discussed above, we have that an infinitesimal transformation generated by
$G_i^1 - \tilde{G}_i^1$ produces the following
boundary terms in the action of the folded theory,
\bea
\delta_{1-\tilde{1}}S &&= \sum_{i=1,2} \int d\tau  \epsilon^i \bigg(r^2_i( \Psi^1_i-\tilde{\Psi}_i^1)\frac{\dot{X}_i^{2}}{r^2_i} - \frac{( \Psi^1_i+\tilde{\Psi}_i^1)}{r^2_i}  (r^2_i X'^{2}_i) \no\\
 && - r^1_i(\Psi^{2\dagger}-\tilde{\Psi}^{2\dagger})\frac{\dot{X}_i^1}{r^1_i}+ \frac{(\Psi^{2\dagger}+ \tilde{\Psi}^{2\dagger})}{r_1^1}r^1_1 X'^{1}_i \bigg)
\eea
where we have taken $\epsilon^i_1=-\epsilon^i_{\tilde{1}}= \epsilon^i$.
The boundary terms would vanish upon imposing the boundary conditions (\ref{bosonicbc}) in case we have $\epsilon^i =\epsilon$ and
\be
\sum_{i=1,2} r^{J}_i(\Psi^I_i- \tilde{\Psi}^I_i) =0, \qquad \frac{\Psi^I_1+ \tilde{\Psi}^I_1}{r^J_1}=\frac{\Psi^I_2+ \tilde{\Psi}^I_2}{r^J_2}
\ee
Here, $J\ne I$. These conditions are precisely the fermionic analog of (\ref{bosonicbc}).
We could repeat this procedure with the other three left- and right-moving pairs of supersymmetric transformations.
Since the same fermionic fields $\Psi^I$ are transformed to different bosonic fields $X^I$, the radii $r^I_i$ of the bosonic
fields $X^I_i$ in each branch are not independent for different $I$.
The simplest choice that would ensure the preservation of one copy of the $\cN=4$ supersymmetry
would be setting all the radii at the different branches to the same value, i.e.
\be
r_i^I=r^J_i=r_i
\ee
The conformal supercharges at each junction would then satisfy the boundary conditions
\be
\sum_{i=1,2} T_i-\tilde{T}_i=0,\qquad \sum_{i=1,2} G^a_i -\tilde{G}^a_i =0
\ee
Note that these boundary conditions are consistent with the structure of the supergravity solutions.
Since the internal moduli of the four-torus are turned off,
all the radii of the torus are treated in the same way and the jump in radius is the same for all of them.
Furthermore, the metric ansatz for the BPS Janus solution has only a manifest $S^2$ factor and, consequently, only $SU(2)$ isometry.
Similarly, the boundary condition (\ref{jboundcon}) preserves only a combination of the $SU(2)_L\times SU(2)_R$
superconformal R-symmetry.

\subsection{Calculation of boundary entropy on the CFT side}

As a starting point we consider the conformal field theory defined by a single  boson  on a circle of radius $R$.
This CFT  has  central charge $c=1$. The interface is defined by a jump of the radius
of the boson across $x=0$, i.e. the boson has radius $R=r_-$ for $x<0$ and radius $R=r_+$ for $x>0$.
The folding trick relates this theory to a CFT of two bosons with radii $r_\pm$ defined on the half-space $x>0$.
The boundary condition on the single boson is given by (\ref{bosonicbc}) setting $i=1,2$ and $I=1$,
 \bea\label{bcfoldb}
 \partial_\tau( \cos \theta \; X^1 -\sin\theta  \; X^2)|_{\sigma=0} = \partial_\sigma( \sin\theta \; X^1 +\sin\theta \; X^2)|_{\sigma=0} =0
 \eea

\begin{figure}
\centering
\includegraphics[scale=0.56]{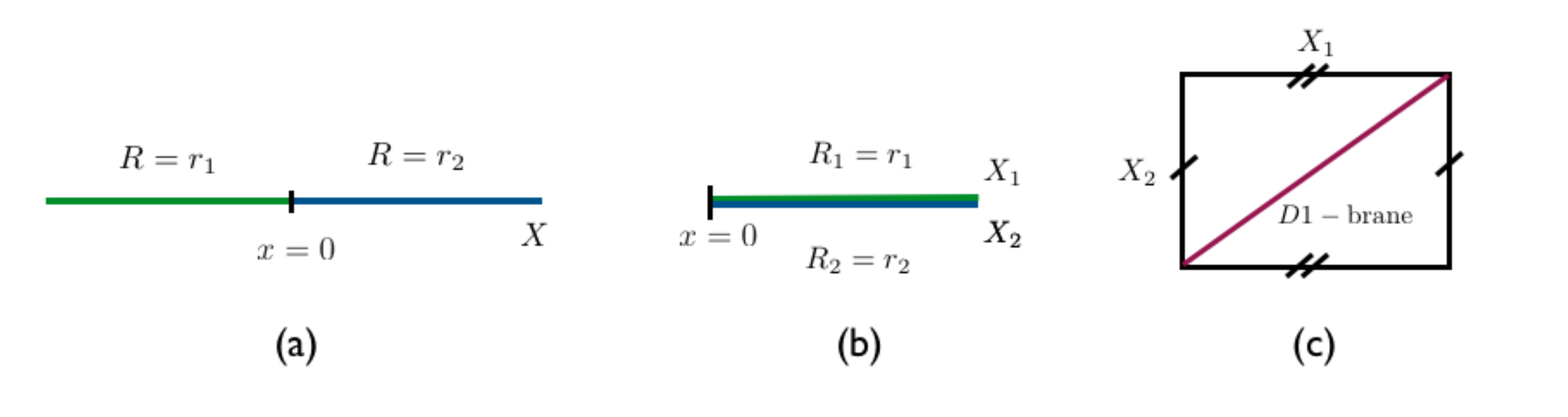}
\caption{(a) An interface where radius of a compact  boson jumps at $x=0$. (b) Folded boundary CFT  and two bosons having  different radiii. (c) Boundary conditions are equivalent to  a diagonal  $D1$ brane in a two dimensional torus. }
\label{fig:1}
\end{figure}

  Hence, after folding, these  conformal interface conditions imply boundary conditions which describe a $D1$-brane on
the diagonal of a rectangular two-torus with radii
 $r_+$ and $r_-$. The boundary entropy of this boundary CFT is defined to be
 \be
 g=\langle 0| B\rangle
 \ee
 The $g$-factor was  calculated in  \cite{Elitzur:1998va} and is given by the tension of the diagonal $D1$-brane on the torus
   \bea\label{gfactor}
 g= {1\over \sqrt{2}  }\sqrt{{r_-\over r_+}+{r_+\over r_-}}
 \eea
 The boundary entropy is then given by the logarithm of the $g$-factor.

 For a tensor product of $n$ bosons with the same jump in radius, the $g$-factor is $g^n$ and hence the boundary
entropy  is given by $S=n \log g$.
Since the $g$-factor is related to the tension of the $D$-brane, the result is also valid for superconformal field theories with fermions.

\subsection{Boundary entropy for non-susy Janus}
As argued in  \cite{Azeyanagi:2007qj}, the sigma-model for the $D1/D5$ CFT is associated with the string frame metric, taking the form
\bea\label{sigmacoup}
\int d^2 z e^{-\phi} G_{\mu\nu}^{s} \partial X^\mu \partial X^\nu
\eea
It follows from the Einstein frame metric (\ref{nonbpsjanusa}) that the string frame metric does not have any factor
of $e^\phi$, hence one identifies the radius on the boundary with the asymptotic value of $e^{-\phi/2}$. Therefore,
\bea
{r_+ \over r_-}  = {\lim_{x\to  \infty} e^{-\phi/2}\over  \lim_{x\to  -\infty} e^{-\phi/2}}= \left( {1+\sqrt{2} \gamma  \over 1-\sqrt{2} \gamma }\right) ^{1\over 2 \sqrt{2}}
\eea
The $D1/D5$ CFT has central charge $c=6 N_1 N_5$ and thus $4 N_1 \times N_5$ bosons.
The solution treats all internal directions equally, so that all the bosons jump by the same amount.
The CFT calculation therefore gives
\bea\label{cftresa}
S_{bdy}= 4 N_1 N_5  \log  \left( {1\over \sqrt{2} } \sqrt{  \left( {1+\sqrt{2} \gamma  \over 1-\sqrt{2} \gamma }\right) ^{1\over 2 \sqrt{2}}+ \left( {1-\sqrt{2} \gamma  \over 1+\sqrt{2} \gamma }\right) ^{1\over 2 \sqrt{2}}}\right)
\eea
Note that (\ref{cftresa}) agrees with the holographic result (\ref{karchresult}) obtained from the non-BPS Janus solution
 up to quadratic order in an expansion around $\gamma=0$. However,
the agreement does not hold for higher-order terms.

\subsection{Boundary entropy for BPS Janus}
As it was argued in Section \ref{susyc}, the supersymmetric boundary conditions imply a jump of all scalars by the same amount.
In this section we repeat the CFT calculation of the boundary entropy using  the value of the radius jumps of the BPS Janus solution
(\ref{bpsmeta}).
Interestingly,  we find complete agreement between the CFT calculation and the holographic calculation.
Transforming the Einstein frame metric (\ref{bpsmeta}) to the string frame metric provides an expression for the radii $r_\pm$
of the compact bosons 
\bea
r_\pm =\lim_{x\to \pm \infty}   \Big( f_3^2 e^{-\phi/2}\Big)^{1\over 2}
\eea
The relevant supergravity solution is the one where $\theta=0$, i.e. there is no jump in the mode dual to the $Z_2$ orbifold   twist operator.
In this case, one finds
\bea
{r_+\over r_-}= e^\psi
\eea
and the boundary entropy for a single boson (\ref{gfactor}) becomes
\bea
g= \sqrt{\cosh \psi}
\eea
Hence, the boundary entropy for the $D1/D5$ CFT becomes
\bea
S_{bdy} &=&  4 N_1 N_5    \log( g)   = {2\over 3} c \log( g)  = {32 \pi^3 L^2 \over k_{10}^2} \cosh^2  \psi \log \cosh ( \psi)
\eea
where we used the formula (\ref{centralc}) for the central charge.
This result agrees exactly with the holographic result obtained in equation (\ref{holores}).

\subsection{Deformation by the $Z_2$ orbifold twist operator}
It is known that the radius deformation of a compact boson is an exactly marginal deformation of the bulk theory,  since the operator
\bea\label{raddeform}
\delta S = \lambda \int d^2z  \;  \partial X \bar \partial X
\eea
is a $U(1)\times U(1)$ current-current deformation \cite{Chaudhuri:1988qb}.
 The case of an interface  located at $x=0$ is more complicated since the deformation is now not uniform,
\be\label{raddeformb}
\delta S = \lambda \int d^2z\;  \Big( \theta(x) \partial X \bar \partial X - \theta(-x)  \partial X \bar \partial X \Big)
\ee
where $\theta(x)$ is the step function. However, it has been shown in  \cite{Fredenhagen:2007rx}
that even in the presence of an interface, the perturbation is exactly marginal and modifies the gluing conditions according
to (\ref{bcfoldb}). It seems likely that this statement can also be directly proven in conformal perturbation theory.

The general holographic result for the boundary entropy contains two deformation parameters, $\theta$ and $\psi$ respectively
\bea\label{holoent}
S_{bdy}&=&  {32 \pi^3 L^2 \over  k_{10}^2}  \cosh^2 \psi \cosh ^2 \theta   \log \big( \cosh \theta \cosh\psi\big) \no \\
&=& 2 N_{D1} N_{D5}  \log \big( \cosh \theta \cosh\psi\big)
\eea
In the previous section we have set $\theta$ to zero. The deformation parameter $\theta$ corresponds to the $Z_2$ orbifold   twist operator.
It is intriguing that the parameter controlling the  radius deformation and the $Z_2$ orbifold twist deformation appear in a completely
symmetric way in the expression (\ref{holoent}).

It is tempting to conjecture that   (\ref{holoent}) is the exact result of a CFT  computation
including both radius and orbifold deformations.

The orbifold deformation is obtained by perturbing the theory by a dimension $(h,\bar h)=(1,1)$ operator ${\cal  T}_0$. The operator
${\cal T}_0$ can be obtained from the following operator product expansion,
\bea\label{defcalt}
{\cal T}_0(z,\bar z) &=& {1\over 2\sqrt{2}}\Big(\epsilon_{ab} \oint dw\; G^{+a}(w) \oint d\bar w \;\bar G^{+b}(\bar  w)  \Sigma^{(1/2,1/2)}(z,\bar z) \no\\
&&+
\epsilon_{ab} \oint dw\; G^{-a}(w) \oint d\bar w \;\bar G^{-b}(\bar  w)  \bar \Sigma^{(1/2,1/2)}(z,\bar z) \Big)
\eea
Here $G^{ab}(z)$ is the superconformal generator and $a$ labels the $SU(2)_R$ charge, whereas $b$ labels the $SU(2)_I$ charge.
Hence, the jump of the modulus associated with the $Z_2$ orbifold twist operator  corresponds to a deformation of the CFT by
\be\label{deformorbi}
\delta S = \lambda \int  d^2z\;\Big( \theta(x)  {\cal T}_0(z,\bar z)  - \theta(-x)   {\cal T}_0(z,\bar z)\Big)
\ee
A first consequence of the result  (\ref{holoent})  is that under an orbifold deformation,
the first non-trivial change in the boundary entropy occurs at second order in $\theta$ and is given
by setting $\psi=0$ and expanding
\bea\label{expansion}
S_{bdy} = {c\over 6} \Big(  \theta^2- {1\over 6} \theta^4+ O(\theta^6)\Big)
\eea
The fact that the term linear in $\theta$ vanishes is in agreement with the result obtained by conformal
perturbation theory in \cite{Green:2007wr}. As discussed in \cite{Green:2007wr}, the change of the $g$-function is given by
\be
{\delta g \over g} = -{\pi \over 2} \theta A_{orb} + O(\theta^2)
\ee
Where $A_{{orb}}$ is the  one-point function of the (unintegrated) operator in (\ref {deformorbi}).
 Since  the one-point function of a single twist operator  evaluated in the unperturbed  bulk theory vanishes,
the lowest nontrivial contribution could appear  at second order, in agreement with  (\ref{expansion}).

Furthermore in Appendix  \ref{appendixb} we show that the operator product expansion of two twist fields is given by
\be
\lim_{|z- w|\to 0} {\cal T}_0(z,\bar z) {\cal T}_0(w,\bar w)= {1\over |z-w|^4} + {\rm finite}
\ee
Where the singular OPE is exactly of the same form as the OPE of the $\partial X\bar \partial X$ operator in the radius deformation (\ref{raddeform}). The
correlation functions on the plane which enter in the conformal perturbation
theory are completely determined by the singular part of the OPE.
Consequently, the radial perturbation (\ref{raddeformb}) and the twist field
perturbation (\ref{deformorbi}) lead to the same change in the boundary entropy.
Note however that due to the complexity  of higher order conformal perturbation theory an explicit higher order calculation has not been performed.

Another view point of the twist  mode deformation may be useful here. The orbifold point is after all one particular point in the full moduli space, and the twist  modes of the orbifold CFT are only some of the many possible K\"ahler moduli deformations. To take a concrete example, consider the orbifold point $T^4/\mathbb{Z}_2$ of a $K_3$ surface. The number of distinct two forms is $b_2= 22$ for a $K_3$ surface but the corresponding number on a $T^4$ is only 6, which can roughly be identified with the world-sheet $B$-fields in the sigma model on $T^4/\mathbb{Z}_2$. The remaining  16 modes arise in the sigma-model precisely as twist-fields deformations. From this perspective then, while the twist-fields have a very different representation in the orbifold CFT, their underlying origin in the moduli space of $K_33$ surfaces are not so fundamentally different from the world-sheet $B$ fields. It is thus tempting to compare the supergravity result with a CFT computation of the boundary entropy in the presence of both  $B$-field and radii jumps on a free CFT, where the jump of the $B$-field is taken to be proportional to that of the twists field couplings read off from the supergravity solution,  i.e. we take $B^\pm =\pm 4 n k \sinh \theta$ where $n$ is a proportionality constant to be determined. Our normalization is such that $2B$ couples to the operator $\partial_+ X^1 \partial_- X^2 - \partial_+ X^2\partial_- X^1$.
The computation is detailed in Appendix C. The result of the computation gives, remarkably for the naive choice of $n= \frac{1}{2}$, precisely $g = \cosh\theta \cosh\psi$, yielding complete agreement with the supergravity theory.

It would be very interesting to perform a direct calculation in the orbifold CFT, perhaps along the lines of  \cite{Fredenhagen:2007rx}  to show  that (\ref{holoent}) is indeed the boundary entropy of a deformed theory including the twist deformation.

\section{Conclusions}\label{sec7}
\setcounter{equation}{0}

In this paper we have used the AdS/CFT correspondence
to calculate holographically the entanglement entropy
for an interface theory given by a marginal deformation of the ${(T^4)}^N / S_N$ orbifold CFT. In this theory,
the volume of the target space four-torus and the  mode dual to the $Z_2$ orbifold twist operator assume different values on each side of the interface.
The calculation was performed using the half-BPS Janus solution obtained in \cite{Chiodaroli:2009yw} as the gravitational
dual for the interface theory.

We found exact agreement between holographic and CFT calculations in the case in which the jump
in the moduli of the conformal field theory corresponds to a change of the six-dimensional dilaton in the supergravity solution.
It appears that the boundary entropy is protected by supersymmetry, even though it is not an index.

The supergravity solution also permits the calculation of the entanglement entropy for a jump of the $Z_2$ orbifold twist operator 
 of the $(T^4)^N/S_N$ orbifold CFT.
It is interesting to note that the expression for the boundary entropy is completely symmetric in the jump
parameters related to the volume and twist  mode deformations.

Since  the marginal operator associated with the twist mode deformation is a twist field (denoted with $ {\cal  T}_0$),
there is no exact conformal field calculation to compare with the supergravity result.
However, it is quite likely that the CFT calculation should agree with the supergravity computation.
This observation leads to some interesting expectations regrading the properties of twist-field correlators.
For a theory with a  compact boson on the plane, the change of radius is related to a deformation
 by an operator of the form $J\bar J$, where $J$ is the R-symmetry current. This operator is exactly marginal  \cite{Chaudhuri:1988qb}.
Furthermore,  correlation functions of an arbitrary number of $J$ operators are obtained from the two-point functions  using Wick's theorem.
We can consider a calculation of the boundary entropy using conformal perturbation theory along the lines of \cite{Green:2007wr}
for both the radius and the twist field deformations.
The result of both calculations are expected to have exactly the same form.
This observation suggests that the correlation functions for the
${\cal  T}_0$ twist fields may be identical to the ones of the $J \bar J$ operators.
 
The study of interface conformal field theories has produced very exciting new developments
with their application to the description of quantum wires (see e.g. \cite{Kane:1992zza,Wong:1994pa,Bellazzini:2008mn}).
Systems of quantum wires display a rich variety of interesting phenomena, with many different IR fixed points,
whose physics is yet to be understood \cite{Oshikawa:2005fh}.
Networks of quantum wires can potentially be engineered using general multi-Janus solutions, and
the entanglement entropy is only one of the many quantities which can be calculated using the AdS/CFT correspondence.
It would be very interesting to obtain the boundary conditions at the interface and  physical observables
such as transport coefficient of the system, which can be extracted using bulk-boundary correlators in the Janus background.

In this paper, we have computed the entanglement entropy only for Janus solutions with $AdS_3$ asymptotics.
However, the framework we have discussed at the beginning of this section can also be applied to BPS Janus
solutions in different dimensions.
In particular, it would be very interesting to obtain the entanglement entropy for the supersymmetric Janus solutions of
\cite{D'Hoker:2007xz} and \cite{D'Hoker:2008qm,D'Hoker:2009gg}.

We plan to return to these interesting topics in the future.

\bigskip\bigskip

\noindent{\Large \bf Acknowledgements}

\medskip
We are grateful to C.~Bachas, E.~D'Hoker, J.~Estes, P.~Kraus, D.~Krym, S.~Mathur and B.~Shieh for useful conversations. The work of MG and MC was
supported in part by NSF grant PHY-07-57702. The work of MC was supported in part by the 2009-10 Siegfried W. Ulmer Dissertation Year Fellowship of UCLA. The research  of LYH at  the Perimeter Institute is supported by the Government of Canada
through Industry Canada and by the Province of Ontario through the Ministry of Research
and Innovation.

\newpage

\appendix

\section{BPS interface solution}\label{appa}
In this appendix we review the regular interface solution of type IIB supergravity
which is locally asymptotic to $AdS_3\times S^3\times M_4$ and preserves eight of the sixteen supersymmetries of the $AdS$ vacuum.
More details can be found in \cite{Chiodaroli:2009yw} and \cite{CGK2010}.

\subsection{Local solutions}\label{appa1}
The solutions are parameterized  by two meromorphic functions $A(z),B(z)$ and two harmonic function $H(z,\bar z),K(z,\bar z)$
(as well as their dual harmonic functions $\tilde K(z,\bar z)$).
All functions depend on the coordinates  of the two-dimensional Riemann surface $\Sigma$ which has a boundary.
The ten-dimensional metric is given by a fibration of $AdS_2\times S^2\times K_3$ over $\Sigma$.
  \be
ds^{2} = f_{1}^{2 } ds^{2}_{AdS_{2}} + f^{2}_{2}ds^{2}_{S^{2}} + f^{2}_{3}ds^{2}_{K_{3}}  + \rho^{2 }dz  d\bar z
\ee
The complex three-form  is
\be
G= g^{(1)}_{a} f_1^2 \;e^{a}\wedge \omega_{AdS_2}+ g^{(2)}_{a} f_2 ^2\; e^{a}\wedge \omega_{S^2}, \quad
\ee
The self-dual five-form flux is given by
\be
F_{5}= h_{a}   f_1^2 f_2^3 \; e^{a}\wedge \omega_{AdS_2}\wedge \omega_{S^2},+ \tilde h_{a} f_3^4  \; e^{a} \wedge \omega_{K_3}
\ee
Dilaton and axion are given by \footnote{In the previous paper we used $\phi$ which is related to the standard dilaton by $\phi=-2\Phi$}
\bea\label{dilformula}
e^{-2\phi} & = & {1\over 4K^2} \Big( (A + \bar A)K  - (B + \bar B)^2 \Big)
		\Big( (A + \bar A)K - (B - \bar B)^2  \Big) \\
 \chi &=& {i\over 2K} \Big( (A - \bar A)K -B^2 +\bar B^2  \Big) \label{axionform}
\eea
 The R-R four-form potential $C_K$ is given by
 \bea\label{cfourexp}
C_K= -{i \over 2} {B^2-\bar B^2\over A+\bar A} -{1\over 2}\tilde K
\eea
and the metric factor of the compact four manifold $M_4$ is given by
\bea\label{f34formula}
f_3^4 &=& 4  {  e^{-\phi} K \over A + \bar A}
\eea
The metric factors associated with $AdS_2$ and $S^2$ are given by
\bea
f^2_1 &=&  {e^{ \phi} \over 2 f_3^2} {|H| \over K}  \Big( (A + \bar A) K  -  (B - \bar B)^2 \Big) \label{sol-f1} \\
f^2_2 &=&  {e^{\phi} \over 2 f_3^2} {|H| \over K} \Big(  (A+ \bar A) K -   (B  + \bar B)^2  \Big) \label{sol-f2} \eea
The rank three anti-symmetric  tensor fields can be expressed in terms of potentials
\bea f_1^2 \rho e^{\phi/2} \Re(g^{(1)})_z & = & \partial_w b^{(1)} \label{potdef1}\\
 f_2^2 \rho e^{\phi/2} \Re(g^{(2)})_z & = & \partial_w b^{(2)} \label{potdef2b}\\
f_1^2 \rho e^{-\phi/2} \Im(g^{(1)})_z + \chi f_1^2 \rho e^{\phi/2} \Re(g^{(1)})_z  & = & \partial_w c^{(1)} \label{potdef3}\\
f_2^2 \rho e^{-\phi/2} \Im(g^{(2)}) _z+ \chi f_2^2 \rho e^{\phi/2} \Re(g^{(2)})_z & = & \partial_w c^{(2)} \label{potdef4}\eea
The potentials $c^{(1,2)}$ and $b^{(1,2)}$ are  expressed in terms of the meromorphic and harmonic functions as follows
\bea b^{(1)} &=& - {H (B + \bar B) \over (A + \bar A) K - (B + \bar B)^2 } - h_1, \qquad h_1={1 \over 2} \int {\partial_w H \over B} + c.c. \label{potharmonic1app}\\
 b^{(2)} &=& -i  {H (B - \bar B) \over (A + \bar A) K - (B - \bar B)^2 } + \tilde h_1, \qquad  \tilde h_1={1 \over 2 i} \int {\partial_w H \over B} + c.c. \label{potharmonic2app}\\
c^{(1)} & = & - i {H (A \bar B -  \bar A B) \over (A + \bar A) K - (B + \bar B)^2 } + \tilde h_2, \qquad \tilde h_2={1 \over 2 i} \int {A \over B}\partial_w H + c.c.  \label{potharmonic3app}\\
c^{(2)} & = & - {H (A \bar B +  \bar A B) \over (A + \bar A) K - (B - \bar B)^2 } + h_2, \qquad  h_2={1 \over 2 } \int {A \over B}\partial_w H + c.c.  \label{potharmonic4app}\eea

\subsection{Page charges}
In this section we review the expressions for the Page charges of the BPS interface. Further details can be found in \cite{CGK2010}.
The Page charges  are  conserved and localized as well as related to the quantized number of branes in a supergravity solution \cite{Marolf:2000cb}.
In type IIB the Page charges for $NS5$ and $D5$-branes are given by
\bea
Q_{NS5}= \int_{M_3} H_3, \quad\quad  Q_{D5} = \int_{M_3} \Big(\tilde F_3 +\chi H_3\Big)\
\eea
The final expressions for the five brane charges are given by
\bea
Q_{NS5}&=&  4\pi \big(\int_{\cal C} dz \; \partial_z b^{(2)} + c.c \big) \no\\
Q_{D5} &=&  4\pi \big(\int_{\cal C} dz \; \partial_z c^{(2)} + c.c \big)
\eea
Where $\cal C$ is a contour in the Riemann surface $\Sigma$ which produces a three-sphere in the asymptotic region
 together with the fibered $S^2$.
For the solutions in this paper, the Riemann surface $\Sigma$ is the half-plane and the contours
 providing homology three-spheres are the ones which enclose a pole of the harmonic function $H$.

The Page charges  for $D1$-branes and fundamental strings are given by
\bea\label{paged1}
 Q_{D1} &=&-\int_{M_7}\Big( e^{\phi}* \tilde F_3 -4 C_4\wedge H_3\Big) \no\\
 Q_{F1} &=& - \int_{M_7} \Big( e^{-\phi} * H_3 -\chi e^\phi *\tilde F_3 +4 C_4\wedge dC_2\Big)
 \eea
 The expressions for the one-brane charges (\ref{paged1})
are more complicated due to the Hodge dual in their definition and the presence of Chern-Simons terms.
The seven-manifold appearing in the above expressions is a product of $M_4$ and a homology three-sphere
obtained from a contour $\cal C$ just as in the case of the five brane charges.

  \medskip

  \noindent The D1-brane charge  is given by
\bea
Q_{D1}&=& 4 \pi  \Big\{  \int_{\cal C}  {4 K \over A+\bar A}  {(A+\bar A)K-(B+\bar B)^2\over (A+\bar A)K -(B-\bar B)^2} i(  \partial_z c^{(1)} -\chi \partial_z b^{(1)} ) dz \no\\
&&- 2\int_{\cal C} \Big( { i } {B^2-\bar B^2\over A+\bar A} +\tilde K\Big)  \partial_{ z} b^{(2)} dz  \Big\}+c.c.
\eea
The fundamental string charge is given by
\bea
Q_{F1}&=& 4\pi  \Big\{ \int_{\cal C} {\Big( (A+\bar A)K-(B+\bar B)^2\Big)^2 \over K(A+\bar A)} i \partial_z b^{(1)} dz +2\Big({i(B^2-\bar B^2)\over A+\bar A} +\tilde K \Big)   \partial_z c^{(2)} dz\no\\
&&- \int_{\cal C}   {4 K \over A+\bar A}  {(A+\bar A)K-(B+\bar B)^2\over (A+\bar A)K -(B-\bar B)^2}    \;   i\chi \Big(   \partial_z c^{(1)} -\chi \partial_z b^{(1)}\Big) dz \Big\}+ c.c.
\eea

\subsection{BPS Janus solution}
The BPS Janus solution is defined on the strip with  coordinates
\bea
w=x+i y, \quad \quad  x\in {}[-\infty,+\infty{} ], \;\; y\in {}[0,\pi{}]
\eea
The meromorphic and harmonic functions are given by
\bea
H &=& -i L  \sinh (w + \psi) + c.c. \\
A &=& i  k^2  { \cosh \theta + \sinh \theta \cosh w \over \sinh w}   \\
B &=& i  k { \cosh (w + \psi) \over \cosh \psi \sinh w} \\
K &=& i  { \cosh \theta - \sinh \theta \cosh w \over \sinh w} + c.c.
\eea
One obtains the  expressions of  Section \ref{sec4} by plugging these functions into the formulae given in Appendix \ref{appa1} .

\section{Operator product of twist fields}\label{appendixb}

It was argued  that for the twist operator ${\cal T}_0$ given in (\ref{defcalt}) the two terms containing $\Sigma^{(1/2,1/2)}$ and $\bar  \Sigma^{(1/2,1/2)}$ are identical (see for example \cite{Gava:2002xb}). It is convenient to evaluate the OPE as follows:

\bea
\lim_{z_1\to  z_2 } {\cal T}_0(z_1,\bar z_1) {\cal T}_0(z_2, \bar z_2) &=&{1\over 2} \lim_{ z_1\to z_2 }   \epsilon_{ab} \Big(\oint_{C_{z_1}} dw \; G^{+a}(w) \oint_{C_{\bar z_1}} d\bar w \; \tilde G^{+b}( \bar w)\Big) \Sigma^{(1/2,1/2)}(z_1,\bar z_1) \no\\
&&\quad \quad \epsilon_{cd} \Big(\oint_{C_{z_2}} dy\; G^{-c}(y) \oint_{C_{\bar z_2}} d\bar y \;\tilde G^{-d}( \bar y)\Big) \bar\Sigma^{(1/2,1/2)}(z_2,\bar z_2) \no\\
\eea
One can now deform the w-contour  $C$  which surrounds $z_1$ in such a way that the contour will surround $z_2$. Since $G^{+a}$ annihilates $\bar \Sigma$ the only relevant term comes from the OPE of two $G$ \cite{Yu:1987dh}:
\bea
\oint_{C_{y}} dw \;G^{++}(w) G^{--}(y) &=& T(y) + \partial_y J^3(y)
 \no\\
   \oint_{C_{y}} dw \;G^{+-}(w) G^{-+}(y)&=&T(y) + \partial_y J^3(y)
\eea
Which can be seen from the fact that the contour $C_y$ surrounds  $y$ and in contour integration  only the simple pole term in the GG OPE survives.
Hence,
\bea\label{ttope}
&&\lim_{z_1\to z_2}{\cal T}_0(z_1,\bar z_1){\cal  T}_0(z_2, \bar z_2)  \no\\
&&=\lim_{z_1\to z_2} \Sigma^{(1/2,1/2)}(z_1,\bar z_1)  \oint_{C_{z_2}} dy  \big(T(y) + \partial_y J^3(y)\big) \oint_{C_{\bar z_2}}d\bar y  \big(T(\bar y) + \bar \partial_y J^3(\bar y)\big) \bar  \Sigma^{(1/2,1/2)}(z_2,\bar z_2)\no\\
\eea
We use  following OPE's

\begin{figure}
\centering
\includegraphics[scale=0.44]{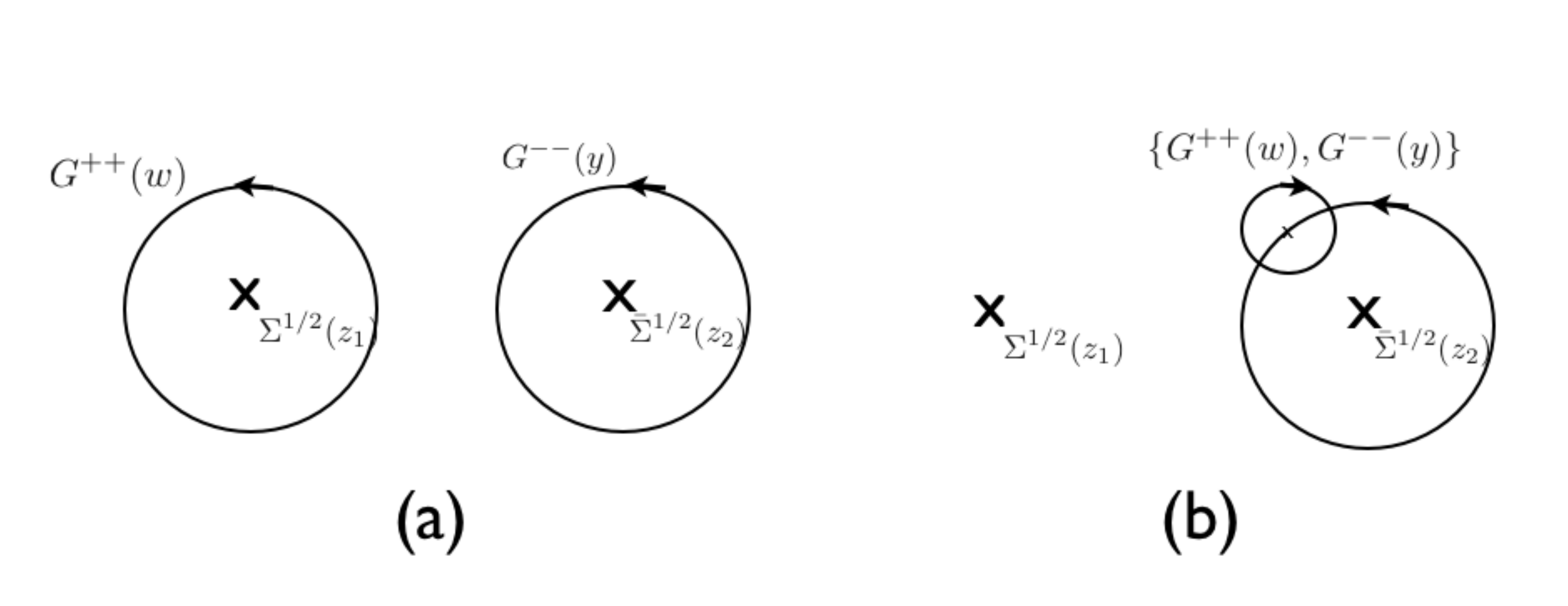}
\caption{(a) Contour for holomorphic part  before deformation (b) Contour deformation picks up anti-commutator of G's. }
\label{fig:contour}
\end{figure}

\bea
T(z) \Sigma^{1/2} (w) &=& {\partial_w \Sigma^{1/2} (w)  \over z-w} +{1\over 2} {\Sigma^{1/2} (w) \over (z-w)^2}+\cdots  \no \\
J^3(z) \Sigma^{1/2} (w) &=& {1\over 2}   {\Sigma^{1/2} (w)\over z-w}
\eea
and hence,
\bea
\Big( T(z)+ \partial_z J^3(z) \Big) \Sigma^{1/2}(w) &=&  {\partial_w \Sigma^{1/2} (w)  \over z-w} +{1\over 2} {\Sigma^{1/2} (w) \over (z-w)^2} + \partial_z {1\over 2}   {\Sigma^{1/2} (w)\over z-w} \no\\
&=&  {\partial_w \Sigma^{1/2} (w)  \over z-w}
\eea
The anti-holomorphic part leads to an analogous
result. Hence the ${\cal T}_0  {\cal T}_0$ OPE (\ref{ttope}) becomes
\bea
\lim_{z_1\to z_2}{\cal T}_0(z_1,\bar z_1) {\cal T}_0(z_2, \bar z_2)   &=&\Sigma^{(1/2,1/2)}(z_1,\bar z_1)  \partial_{z_2} \partial_{\bar z_2}  \bar \Sigma^{(1/2,1/2)}(z_2,\bar z_2) \no \\
&=&  \partial_{z_2} \partial_{\bar z_2}   {1\over (z_1-z_2) (\bar z_1 -\bar z_2)}  +\cdots \no\\
&=& {1\over |z_1-z_2|^4} + \cdots
\eea
Where the dots denote nonsingular terms as $z_1\to z_2$.
We used the fact that the OPE of the chiral and antichiral twist field are given by
\be
\lim_{z_1\to z_2} \Sigma^{(1/2,1/2)}\bar \Sigma^{(1/2,1/2)} = {1\over (z_1-z_2)(\bar z_1 -\bar z_2)} + \cdots
\ee

\section{Boundary entropy with both radii and $B$ field jumps}

Consider starting off with two compact bosons from each side of the interface, which is the smallest number of fields that can be  coupled to an antisymmetric $B$-field.

The action before folding is
\begin{equation}
\int_{\sigma>0} d\sigma d\tau \big(r_i^2 \eta^{ab} \partial_a X^i\partial_b X^i + 2\epsilon^{ab} b\partial_a X^1\partial_b X^2 \big)+ \int_{\sigma<0} d\sigma d\tau \big(\tilde{r}_i^2 \eta^{ab} \partial_a \tilde{X}^i\partial_b \tilde{X}^i + 2\epsilon^{ab} \tilde{b}\partial_a \tilde{X}^1\partial_b \tilde{X}^2\big)
\end{equation}
where $i \in \{1,2\}$, and $a,b\in\{\sigma,\tau\}$. After folding the theory in $\sigma<0$, by sending $\sigma\to-\sigma$, we obtain the same action except $\tilde{b}$ changes sign.
The boundary conditions following from the vanishing of the boundary terms upon variation of the fields are given by
\begin{eqnarray}\label{boundcona}
&&r^2_i\partial_\sigma X^i + \tilde{r}^2_i\partial_\sigma \tilde{X}^i + \epsilon_{ij} (b  \partial_\tau X^j - \tilde{b} \partial_\tau \tilde{X}^j) =0, \quad\quad i=1,2\no\\
&&\partial_\tau X^i = \partial_\tau \tilde{X}^i, \quad\quad i=1,2
\end{eqnarray}
The boundary conditions (\ref{boundcona}) correspond to the  \emph{open strings} picture, where boundary conditions are imposed on the spatial boundary $\sigma=0$ of the world-sheet. By exchanging $\sigma \leftrightarrow \tau$ one obtains  the \emph{closed string} picture, where a closed string propagates and the boundary conditions are imposed at a fixed time $\tau=0$.

The mode expansion of the closed strings are given by (using the normalization in \cite{Elitzur:1998va})
\begin{equation}
X^i(\sigma_\tau) = x^i + 2 \omega_i \sigma + \tau G^{ij}(p_j-2B_{jk}\omega^k) + \frac{i}{2}\sum_{n\ne 0}\frac{1}{n} \left( \alpha^i_{L,n} e^{-2in\sigma_+} + \alpha^i_{R,n} e^{-2in\sigma_-}\right)
\end{equation}
where $p_i$ and $\omega^i$ are the momentum and winding respectively and in our  normalization   these are integers. The target space metric $G_{ij}$ is again simply $\delta_{ij} r_i^2$, whereas $B_{ij} = b$. We have similar expressions for   the tilde variables, except the sign of $\tilde{b}$ is flipped. Note that the position of the indices are important.

Using the above expansion, the boundary conditions (\ref{boundcona}) implies here
\begin{equation}\label{cbc}
p_i =-\tilde{p}_i, \qquad \omega^i = \tilde{\omega}^i
\end{equation}
These relations dictate how we build the corresponding boundary state $|B\rangle $. However, to extract the boundary  entropy $\langle 0|B\rangle$ we need only to know the zero modes contribution of $\langle B|\exp(-i\pi H/T)|B\rangle $, where $H$ is the closed string Hamiltonian and $T$ the periodicity in the time direction.
The Hamiltonian is given by
\begin{equation}
H= \frac{1}{4} \bigg(p_\mu G^{\mu\nu}p_\nu + 4\omega^\mu(G-B\cdot G^{-1}\cdot B)_{\mu\nu}\omega^\nu + 4\omega^\mu(B\cdot G^{-1})_\mu^\nu p_\nu \bigg) + N_L + N_R
\end{equation}
where indices $\mu,\nu$ denote summation over both tilded and untilded fields.  $N_{L,R}$ are the level of the left and right moving modes.

Using (\ref{cbc}), we can rewrite the zero-mode contribution of the Hamiltonian as a matrix multiplication
$P^T\cdot M\cdot P$,where
\begin{equation}
P = \left(\begin{array}{c}p_1\\p_2\\\omega_1\\\omega_2\end{array}\right), \qquad M=\frac{1}{4}\left(\begin{array}{cccc}M_{11}&0&0&M_{14}\\
0&M_{22}&M_{23}&0\\
0&M_{32}&M_{33}&0\\
M_{41}&0&0&M_{44}\end{array}\right)
\end{equation}
and $M$ is a symmetric matrix where
\begin{eqnarray}
&&M_{11}= \frac{1}{r_1^2}+ \frac{1}{\tilde{r}_1^2},\qquad M_{22}= \frac{1}{r_2^2}+ \frac{1}{\tilde{r}_2^2}\nonumber \\
&&M_{33}= 4\left(r_1^2 + \frac{b^2}{r_2^2}+\tilde{r}_1^2 + \frac{\tilde{b}^2}{\tilde{r}_2^2}\right),\qquad M_{44}= 4\left(r_2^2 + \frac{b^2}{r_1^2}+\tilde{r}_2^2 + \frac{\tilde{b}^2}{\tilde{r}_1^2}\right)\nonumber \\
&&M_{14}= -2\left(\frac{b}{r_1^2}+\frac{\tilde{b}}{r_1^2}\right),\qquad M_{23}= 2\left(\frac{b}{r_2^2}+\frac{\tilde{b}}{r_2^2}\right)
\end{eqnarray}
Th boundary entropy $g_b$ is then given by
\begin{equation}
g_b= |\det M|^{1/4} = \bigg\vert\frac{((b -\tilde{b})^2 + (r_1^2 + \tilde{r}_1^2) (r_2^2 +
     \tilde{r}_2^2))^2}{16 r_1^2 r_2^2 \tilde{r}_1^2 \tilde{r}_2^2}\bigg\vert^{1/4}
\end{equation}
Now using (\ref{twistcoup},\ref{radii}), we take
\begin{equation}
b= -\tilde{b}=4nk \sinh\theta ,\qquad r_i = \sqrt{\frac{2 k e^{\psi}}{\cosh\psi}},\qquad \tilde{r}_i = \sqrt{\frac{2 k e^{-\psi}}{\cosh\psi}}
\end{equation}
For $n=1/2$
\begin{equation}
g_b= \cosh\theta \cosh\psi
\end{equation}
Since we are considering two directions along $T^4$ at the same time in the above calculation, the final result, including the other two orthogonal directions in $T^4$ and all the copies in the symmetric product,
is indeed
\begin{equation}
S_{bdy} = 2 N_{D1}N_{D5}\log(\cosh\theta \cosh\psi)
\end{equation}
in complete agreement with the supergravity result (\ref{holoent}).


\newpage

\begin{thebibliography}{99}



\bibitem{Cardy:1989ir}
  J.~L.~Cardy,
  ``Boundary Conditions, Fusion Rules And The Verlinde Formula,''
  Nucl.\ Phys.\  B {\bf 324} (1989) 581.

\bibitem{Oshikawa:1996dj}
  M.~Oshikawa and I.~Affleck,
  ``Boundary conformal field theory approach to the critical  two-dimensional
  Ising model with a defect line,''
  Nucl.\ Phys.\  B {\bf 495} (1997) 533
  [arXiv:cond-mat/9612187].



\bibitem{Bachas:2001vj}
  C.~Bachas, J.~de Boer, R.~Dijkgraaf and H.~Ooguri,
  ``Permeable conformal walls and holography,''
  JHEP {\bf 0206}, 027 (2002)
  [arXiv:hep-th/0111210].


\bibitem{Affleck:1991tk}
  I.~Affleck and A.~W.~W.~Ludwig,
  ``Universal noninteger 'ground state degeneracy' in critical quantum
  systems,''
  Phys.\ Rev.\ Lett.\  {\bf 67} (1991) 161.


\bibitem{PhysRevLett.67.2882}
 Seaman, C. L. and Maple, M. B. and Lee, B. W. and Ghamaty, S.  and Torikachvili, M. S. and Kang, J.-S.  and Liu, L. Z. and Allen, J. W. and Cox, D. L.,
"Evidence for non-Fermi liquid behavior in the Kondo alloy $Y1-x$$Ux$$Pd3$,''
Phys.\ Rev.\ Lett.\ {\bf 67} (1991) 2882.




\bibitem{Calabrese:2004eu}
  P.~Calabrese and J.~L.~Cardy,
  ``Entanglement entropy and quantum field theory,''
  J.\ Stat.\ Mech.\  {\bf 0406} (2004) P002
  [arXiv:hep-th/0405152].

\bibitem{Maldacena:1997re}
  J.~M.~Maldacena,
  ``The large N limit of superconformal field theories and supergravity,''
  Adv.\ Theor.\ Math.\ Phys.\  {\bf 2} (1998) 231
  [Int.\ J.\ Theor.\ Phys.\  {\bf 38} (1999) 1113]
  [arXiv:hep-th/9711200].

\bibitem{Gubser:1998bc}
  S.~S.~Gubser, I.~R.~Klebanov and A.~M.~Polyakov,
  ``Gauge theory correlators from non-critical string theory,''
  Phys.\ Lett.\  B {\bf 428} (1998) 105
  [arXiv:hep-th/9802109].

\bibitem{Witten:1998qj}
  E.~Witten,
  ``Anti-de Sitter space and holography,''
  Adv.\ Theor.\ Math.\ Phys.\  {\bf 2} (1998) 253
  [arXiv:hep-th/9802150].





\bibitem{Karch:2000gx}
  A.~Karch and L.~Randall,
  ``Open and closed string interpretation of SUSY CFT's on branes with
  boundaries,''
  JHEP {\bf 0106}, 063 (2001)
  [arXiv:hep-th/0105132].




\bibitem{Aharony:2003qf}
  O.~Aharony, O.~DeWolfe, D.~Z.~Freedman and A.~Karch,
  ``Defect conformal field theory and locally localized gravity,''
  JHEP {\bf 0307} (2003) 030
  [arXiv:hep-th/0303249].



\bibitem{Bak:2003jk}
  D.~Bak, M.~Gutperle and S.~Hirano,
  ``A dilatonic deformation of AdS(5) and its field theory dual,''
  JHEP {\bf 0305} (2003) 072
  [arXiv:hep-th/0304129].







\bibitem{Clark:2005te}
  A.~Clark and A.~Karch,
  ``Super Janus,''
  JHEP {\bf 0510}, 094 (2005)
  [arXiv:hep-th/0506265].



\bibitem{Lunin:2006xr}
  O.~Lunin,
  ``On gravitational description of Wilson lines,''
  JHEP {\bf 0606}, 026 (2006)
  [arXiv:hep-th/0604133].


\bibitem{Lunin:2007ab}
  O.~Lunin,
  ``1/2-BPS states in M theory and defects in the dual CFTs,''
  JHEP {\bf 0710}, 014 (2007)
  [arXiv:0704.3442 [hep-th]].

\bibitem{Yamaguchi:2006te}
  S.~Yamaguchi,
  ``Bubbling geometries for half BPS Wilson lines,''
  Int.\ J.\ Mod.\ Phys.\  A {\bf 22} (2007) 1353
  [arXiv:hep-th/0601089].

\bibitem{Gomis:2006sb}
  J.~Gomis and F.~Passerini,
  ``Holographic Wilson loops,''
  JHEP {\bf 0608} (2006) 074
  [arXiv:hep-th/0604007].


\bibitem{Gomis:2006cu}
  J.~Gomis and C.~Romelsberger,
  ``Bubbling defect CFT's,''
  JHEP {\bf 0608} (2006) 050
  [arXiv:hep-th/0604155].

\bibitem{D'Hoker:2007xy}
  E.~D'Hoker, J.~Estes and M.~Gutperle,
  ``Exact half-BPS Type IIB interface solutions I: Local solution and
  supersymmetric Janus,''
  JHEP {\bf 0706} (2007) 021
  [arXiv:0705.0022 [hep-th]].


\bibitem{D'Hoker:2007xz}
  E.~D'Hoker, J.~Estes and M.~Gutperle,
  ``Exact half-BPS type IIB interface solutions. II: Flux solutions and
  multi-janus,''
  JHEP {\bf 0706} (2007) 022
  [arXiv:0705.0024 [hep-th]].



\bibitem{D'Hoker:2008wc}
  E.~D'Hoker, J.~Estes, M.~Gutperle and D.~Krym,
  ``Exact Half-BPS Flux Solutions in M-theory I, Local Solutions,''
  JHEP {\bf 0808} (2008) 028
  [arXiv:0806.0605 [hep-th]].

\bibitem{D'Hoker:2008qm}
  E.~D'Hoker, J.~Estes, M.~Gutperle and D.~Krym,
  ``Exact Half-BPS Flux Solutions in M-theory II: Global solutions asymptotic
  to $AdS_7 \times  S^4$,''
  JHEP {\bf 0812} (2008) 044
  [arXiv:0810.4647 [hep-th]].

\bibitem{D'Hoker:2009my}
  E.~D'Hoker, J.~Estes, M.~Gutperle and D.~Krym,
  ``Exact Half-BPS Flux Solutions in M-theory III: Existence and rigidity of
  global solutions asymptotic to $AdS4 \times S7$,''
  arXiv:0906.0596 [hep-th].


\bibitem{Chiodaroli:2009yw}
  M.~Chiodaroli, M.~Gutperle and D.~Krym,
  ``Half-BPS Solutions locally asymptotic to $AdS_3 \times  S^3$ and interface
  conformal field theories,''
  JHEP {\bf 1002}, 066 (2010)
  [arXiv:0910.0466 [hep-th]].

\bibitem{Kumar:2002wc}
  J.~Kumar and A.~Rajaraman,
  ``New supergravity solutions for branes in  $AdS_3 \times  S^3$,''
  Phys.\ Rev.\  D {\bf 67} (2003) 125005
  [arXiv:hep-th/0212145].

\bibitem{Kumar:2003xi}
  J.~Kumar and A.~Rajaraman,
  ``Supergravity solutions for  $AdS_3\times   S^3$ branes,''
  Phys.\ Rev.\  D {\bf 69} (2004) 105023
  [arXiv:hep-th/0310056].

\bibitem{Kumar:2004me}
  J.~Kumar and A.~Rajaraman,
  ``Revisiting D-branes in $AdS_3 \times   S^3$,''
  Phys.\ Rev.\  D {\bf 70} (2004) 105002
  [arXiv:hep-th/0405024].

\bibitem{Ryu:2006bv}
  S.~Ryu and T.~Takayanagi,
  ``Holographic derivation of entanglement entropy from AdS/CFT,''
  Phys.\ Rev.\ Lett.\  {\bf 96} (2006) 181602
  [arXiv:hep-th/0603001].


\bibitem{Ryu:2006ef}
  S.~Ryu and T.~Takayanagi,
  ``Aspects of holographic entanglement entropy,''
  JHEP {\bf 0608} (2006) 045
  [arXiv:hep-th/0605073].

\bibitem{Azeyanagi:2007qj}
  T.~Azeyanagi, A.~Karch, T.~Takayanagi and E.~G.~Thompson,
  ``Holographic Calculation of Boundary Entropy,''
  JHEP {\bf 0803} (2008) 054
  [arXiv:0712.1850 [hep-th]].

\bibitem{Bak:2007jm}
  D.~Bak, M.~Gutperle and S.~Hirano,
  ``Three dimensional Janus and time-dependent black holes,''
  JHEP {\bf 0702} (2007) 068
  [arXiv:hep-th/0701108].



\bibitem{CGK2010}
M.~Chiodaroli, M.~Gutperle. L-Y. ~Hung and D.~Krym, "String Junctions and Holographic Interface Solutions in $AdS_3 \times S_3 \times K_3$", to appear.

\bibitem{Papadimitriou:2004rz}
  I.~Papadimitriou and K.~Skenderis,
  ``Correlation functions in holographic RG flows,''
  JHEP {\bf 0410} (2004) 075
  [arXiv:hep-th/0407071].

\bibitem{Brown:1986nw}
  J.~D.~Brown and M.~Henneaux,
  ``Central Charges in the Canonical Realization of Asymptotic Symmetries: An
  Example from Three-Dimensional Gravity,''
  Commun.\ Math.\ Phys.\  {\bf 104} (1986) 207.



\bibitem{Elitzur:1998va}
  S.~Elitzur, E.~Rabinovici and G.~Sarkissian,
  ``On least action D-branes,''
  Nucl.\ Phys.\  B {\bf 541} (1999) 246
  [arXiv:hep-th/9807161].

\bibitem{Chaudhuri:1988qb}
  S.~Chaudhuri and J.~A.~Schwartz,
  ``A CRITERION FOR INTEGRABLY MARGINAL OPERATORS,''
  Phys.\ Lett.\  B {\bf 219} (1989) 291.

\bibitem{Fredenhagen:2007rx}
  S.~Fredenhagen, M.~R.~Gaberdiel and C.~A.~Keller,
  ``Symmetries of perturbed conformal field theories,''
  J.\ Phys.\ A  {\bf 40} (2007) 13685
  [arXiv:0707.2511 [hep-th]].

\bibitem{Green:2007wr}
  D.~R.~Green, M.~Mulligan and D.~Starr,
  ``Boundary Entropy Can Increase Under Bulk RG Flow,''
  Nucl.\ Phys.\  B {\bf 798}, 491 (2008)
  [arXiv:0710.4348 [hep-th]].

\bibitem{Marolf:2000cb}
  D.~Marolf,
  ``Chern-Simons terms and the three notions of charge,''
  arXiv:hep-th/0006117.







\bibitem{David:2002wn}
  J.~R.~David, G.~Mandal and S.~R.~Wadia,
  ``Microscopic formulation of black holes in string theory,''
  Phys.\ Rept.\  {\bf 369} (2002) 549
  [arXiv:hep-th/0203048].


\bibitem{Kane:1992zza}
  C.~L.~Kane and M.~P.~A.~Fisher,
  ``Transmission through barriers and resonant tunneling in an interacting
  one-dimensional electron gas,''
  Phys.\ Rev.\  B {\bf 46} (1992) 15233.


\bibitem{Wong:1994pa}
  E.~Wong and I.~Affleck,
  ``Tunneling in quantum wires: A Boundary conformal field theory approach,''
  Nucl.\ Phys.\  B {\bf 417} (1994) 403.

\bibitem{Bellazzini:2008mn}
  B.~Bellazzini, M.~Burrello, M.~Mintchev and P.~Sorba,
  ``Quantum Field Theory on Star Graphs,''
  arXiv:0801.2852 [hep-th].


\bibitem{Oshikawa:2005fh}
  M.~Oshikawa, C.~Chamon and I.~Affleck,
  ``Junctions of three quantum wires,''
  J.\ Stat.\ Mech.\  {\bf 0602} (2006) P008
  [arXiv:cond-mat/0509675].

\bibitem{D'Hoker:2009gg}
  E.~D'Hoker, J.~Estes, M.~Gutperle and D.~Krym,
  ``Janus solutions in M-theory,''
  JHEP {\bf 0906} (2009) 018
  [arXiv:0904.3313 [hep-th]].


\bibitem{Gava:2002xb}
  E.~Gava and K.~S.~Narain,
  ``Proving the pp-wave / CFT(2) duality,''
  JHEP {\bf 0212} (2002) 023
  [arXiv:hep-th/0208081].

\bibitem{Yu:1987dh}
  M.~Yu,
  ``The unitary  representations  of  the  N=4 SU(2) extended superconformal
  algebras,''
  Nucl.\ Phys.\  B {\bf 294} (1987) 890.

\end{thebibliography}
\end{document}